\documentclass[11pt,a4paper]{article}
\usepackage{jheppub}
\usepackage{amsmath}

\usepackage{amssymb}
\usepackage{bbold}
\usepackage{bm}
\usepackage{color,graphicx}
\usepackage{slashed}
\usepackage{comment, booktabs}
\usepackage{hyperref}
\usepackage{subcaption}
\usepackage{makecell,multirow}
\usepackage[dvipsnames]{xcolor}
\usepackage{pifont}
\usepackage{float}
\usepackage{array}
    \newcolumntype{P}[1]{>{\centering\arraybackslash}p{#1}}
    \newcolumntype{M}[1]{>{\centering\arraybackslash}m{#1}}

\usepackage{hyperref}
\usepackage{tablefootnote}
\usepackage{tabularx}
\definecolor{nicered}{rgb}{0.5,0.,0.}
\definecolor{nicegreen}{rgb}{0.,0.5,0.}
\definecolor{niceblue}{rgb}{0.,0.,0.5}
\hypersetup{colorlinks,citecolor=nicegreen,linkcolor=nicered,urlcolor=niceblue}

\preprint{ANL-198026, MSUHEP-25-017}

\title{QED-enhanced PDF implications for the Higgs sector}

\author[a]{Amanda~M.~Cooper-Sarkar,}
\author[b]{Thomas Cridge,}
\emailAdd{thomas.cridge@uantwerpen.be}
\author[c]{T.~J.~Hobbs,}
\emailAdd{tim@anl.gov}
\author[d]{Joey Huston,}
\author[d]{Pavel Nadolsky,}
\author[d]{Maximiliano Ponce-Chavez,}
\emailAdd{poncedan@msu.edu}
\author[d]{and Keping Xie}

\affiliation[a]{Department of Physics, University of Oxford, OX1 3RH, Oxford, UK}
\affiliation[b]{Elementary Particle Physics, University of Antwerp,\\Groenenborgerlaan 171, 2020 Antwerp, Belgium}
\affiliation[c]{High Energy Physics Division, Argonne National Laboratory, Lemont, IL 60439, USA}
\affiliation[d]{Department of Physics and Astronomy, Michigan State University, East Lansing, MI 48824, USA}

\abstract{
In this work, we examine the implications of electroweak corrections beyond leading order
for processes of special interest in the Higgs sector. We especially explore the role of
these corrections given the introduction of an explicit parton distribution function (PDF) for the photon in the proton,
an object which emerges necessarily in global PDF fits which include QED effects ({\it i.e., QED-enhanced PDFs}). We concentrate on several representative cases, including total
Higgs-production cross sections through gluon fusion, $gg \to H$, vector-boson fusion (VBFH), and associated production, $pp \to VH$; we also examine differential distributions, taking
a representative Higgs-strahlung process, $pp \to W^+H$. We find that the recently developed
LUX formalism for the photon PDF significantly stabilizes the PDF dependence of
both QED-PDF and electroweak corrections in the Higgs sector, while leaving overall $\sim\!3\!-\!4\%$
cross-section-level variations, depending on the chosen QED-enhanced PDF.
We illustrate this QED-enhanced PDF dependence by exploring predictions based upon recent analyses of the CTEQ-TEA, MSHT, and NNPDF analysis groups, fitted either at NNLO or approximate N3LO in QCD.
}
\date{\today}
\begin{document}
\maketitle

%
%
\section{Introduction}
\label{sec:intro}
Recent years have seen significant progress in testing the Standard Model (SM)
at the Energy Frontier. These tests depend critically on high-precision
measurements at the Large Hadron Collider (LHC) and will be advanced
further with heightened statistics in the era of the high-luminosity
LHC (HL-LHC). In this context, the stringency of SM tests and reach
of searches for physics beyond the SM (BSM) require corresponding reductions
in theoretical uncertainties with increasing experimental precision. In particular,
both precision SM and BSM-sensitive electroweak measurements depend on the perturbative
accuracy and the robustness of PDFs which are a necessary input to LHC predictions.
These needs have motivated efforts to develop both the perturbative and
numerical accuracy of calculations for processes at the LHC, including
the calculation of partonic hard scattering to higher orders in both QCD
and electroweak theory.

While multi-loop QCD corrections at and beyond next-to-next-to-leading order (NNLO) accuracy have received substantial recent
attention in the PDF community~\cite{McGowan:2022nag,Jing:2023isu,Cridge:2023ozx,Andersen:2024czj,NNPDF:2024nan,Cooper-Sarkar:2024crx,MSHT:2024tdn},
electroweak effects, including QED corrections, have a subtle interplay with QCD
effects~\cite{Cridge:2023ryv,Barontini:2024dyb,MSHT:2024tdn} and can be essential for refining
predictions of processes like Higgs production and associated production of the
Higgs with weak bosons, so-called Higgs-strahlung.
In addition to modifications to the detailed running of short-distance matrix elements
through mixing with QCD evolution, higher-order QED corrections dynamically generate a
photon PDF~\cite{Manohar:2016nzj,Manohar:2017eqh,Xie:2021equ,Xie:2023qbn,Cridge:2021pxm,NNPDF:2024djq,Cridge:2023ryv,Barontini:2024eii}
starting at NLO in QED. The presence of the photon PDF alters a range of Higgs-production
cross sections, especially those with more pronounced sensitivity to initial-state QED
radiation, including Higgs-strahlung ($VH$) and vector-boson fusion (VBFH).
In addition to their consequential and direct role in electroweak phenomenology, QED
corrections also influence the accuracy, stability, and precision of
parton distribution functions (PDFs)~\cite{Amoroso:2022eow} as extracted from
hadronic data.
This latter aspect motivates the present study.
We revisit several representative LHC processes and assess the impact of effects arising from inclusion of a photon PDF and related
QED corrections to PDFs. We also examine NLO  electroweak corrections, particularly for the representative case of Higgs-strahlung,
$pp\to W^+H$.
We therefore stress that our study concerns {\bf two main effects}, which are distinct but nevertheless related: ({\it i}) the impact of including QED effects (resulting in a photon density) in PDF global fits and the resulting downstream shifts in Higgs-production cross sections; and ({\it ii}) the interplay of these PDF-driven effects with the full NLO electroweak corrections which we compute for a specific case given its connection to photon-initiated contributions, namely, $pp\to W^+H$.
Throughout, we pay special attention to the size and possible
kinematical behavior of these corrections, as well as the dependence on assumed
PDFs extracted at various perturbative accuracies.
We also regard our study as an extension of recent and ongoing
efforts~\cite{PDF4LHCWorkingGroup:2022cjn,Cridge:2021qjj,Jing:2023isu} to benchmark
PDFs for high-energy phenomenology, as we undertake side-by-side comparisons
of predicted cross sections against a range of available QED-enhanced PDFs
determined with various theoretical assumptions.

The remainder of our paper is as follows: after briefly recalling the LUX
formalism in Sec.~\ref{sec:theory}, we provide a concise survey of the PDF-level
impacts of QED corrections in Sec.~\ref{sec:PDFs}. Sec.~\ref{sec:results} contains our main results on the phenomenological impacts of the QED PDFs on various Higgs-production cross sections, including total cross sections (Sec.~\ref{sec:total}) for a number of processes as well as differential distributions for Higgs-strahlung (Sec.~\ref{sec:HV}), for which we evaluate full NLO electroweak corrections; we also briefly summarize our recommendations for phenomenological calculations in the Higgs sector in Sec.~\ref{sec:PDF4LHC}, before concluding in Sec.~\ref{sec:conc}. We include an additional collection of parton-parton luminosity plots in App.~\ref{app:lumis}.

%
%
%
%
\section{QED corrections in the context of electroweak theory}
\label{sec:theory}

The LUX formalism~\cite{Manohar:2016nzj,Manohar:2017eqh} represented a significant step
in electroweak phenomenology. In this approach, the photon PDF of the proton (or of other
hadrons~\cite{Harland-Lang:2019pla,Xie:2023qbn,Cridge:2021pxm}) can be directly related
to knowledge of the $F_{2,L}(x,Q^2)$ structure functions measured in deep-inelastic
scattering (DIS).
The $x$ dependence of the photon PDF at an energy scale, $\mu$, is given by
\begin{align} \label{eq:LUXQED}
x \gamma (x, \mu^2) &= \frac{1}{2\pi \alpha(\mu^2)} \int_x^1 \frac{\mathrm{d}z}{z}
\Bigg\{ \int_{x^2 m_p^2 / (1-z)}^{\mu^2} \frac{\mathrm{d}Q^2}{Q^2} \alpha_{\text{ph}}(-Q^2)
\Bigg[ \Big( z p_{\gamma q}(z) + \frac{2x^2 m_p^2}{Q^2} \Big) 
F_2 \Big(\frac{x}{z}, Q^2 \Big) \nonumber \\
&- z^2 F_L \Big(\frac{x}{z}, Q^2 \Big) \Bigg] 
- \alpha(\mu^2) z^2 F_2 \Big(\frac{x}{z}, \mu^2 \Big)
\Bigg\} + \mathcal{O}(\alpha^2, \alpha \alpha_s)\ ,
\end{align}
and we refer interested readers to, {\it e.g.}, Ref.~\cite{Manohar:2017eqh} for a formal derivation
of Eq.~(\ref{eq:LUXQED}) above as well as exhaustive definitions of quantities appearing therein like
$\alpha_{\text{ph}}(-Q^2)$ and $p_{\gamma q}(z)$. We stress that the practical evaluation of the $Q^2$ moments
in Eq.~(\ref{eq:LUXQED}) requires control over contributions to the structure functions
from low-energy processes below the DIS continuum region, which represent an important source of uncertainty in the photon PDF calculation.
These various contributions can be incorporated systematically alongside corresponding
uncertainty estimates, allowing a new generation~\cite{Xie:2021equ,Cridge:2021pxm,NNPDF:2024djq,Cridge:2023ryv,Barontini:2024dyb}
of QED-enhanced PDF analyses.
In Sec.~\ref{sec:results}, we compare predicted cross sections up to N3LO QCD accuracy (in the case
of total hadronic as well as hard-scattering cross sections), and/or including NLO electroweak corrections (for Higgs-strahlung) assuming
contemporary PDFs including the photon distribution according to the LUX formalism.

%
%

\section{Impact of QED corrections on the PDFs}
\label{sec:PDFs}

The addition of QED DGLAP effects at $\mathcal{O}(\alpha,\alpha^2,\alpha\alpha_S)$ necessarily results in the inclusion of a photon PDF, {\it i.e.}, a small contribution to the partonic content of the proton from the photon. This takes a $\lesssim\! 1\%$ share of the proton's total momentum and therefore affects the distribution of the proton total momentum among the gluon, quark-singlet and photon PDFs. While all global PDF fitting groups now follow the LUXQED approach, there are subtle differences in its application that may alter the precise details of the impacts of the QED corrections on the PDFs. Examples of such methodological variations include:
\begin{itemize}
    \item The scale at which the photon PDF is extracted using equation~\ref{eq:LUXQED} and modified versions thereof. This may also have impacts on the full QCD$\otimes$QED evolution.
    \item The way in which the momentum required by the newly-incorporated photon PDF is extracted from the other partons; {\it i.e.}, is momentum conservation strictly imposed and if so are the details of the redistribution of the proton momentum from the QCD partons determined by the fit or imposed in some other manner. 
    \item The inclusion or otherwise of electroweak corrections in the default QCD baseline upon which the QED corrections are applied.
\end{itemize}
In addition, the impact of QED corrections depends on various precise details of the baseline QCD-only fits, 
such as the distribution of the proton's momentum among the gluon and quark degrees of freedom, data sets included in the PDF fits, and the form of the QCD evolution (exact {\it vs.}~truncated, see, {\it e.g.}, \cite{NNPDF:2024djq,Cooper-Sarkar:2024crx}).

In order to address this, in this work we study a variety of QED-enhanced PDF sets. For comparison with the CT18 NNLO QCD  PDFs~\cite{Hou:2019efy}, we consider CT18$_{\rm QEDfit}$, CT18$_{\rm QEDproton}$, and CT18$_{\rm LUX}$~\cite{Xie:2021equ,Xie:2023qbn} --- the PDF ensembles realizing three possible approaches to the inclusion of QED corrections. In the latter of these, the LUXQED formula for the photon PDF [Eq.~(\ref{eq:LUXQED})] is applied directly, with no QCD$\otimes$QED PDF evolution utilized, therefore there is a small resulting excess in the total momentum of the proton. The former two avoid this by applying the LUXQED formula at a low scale to determine the inelastic photon PDF and then evolving with gluon and quark PDFs consistently in QCD$\otimes$QED evolution to the required scale. In CT18$_{\rm QEDproton}$, the photon momentum is taken from the quark sea by hand, while the CT18$_{\rm QEDfit}$ case allows the fit to determine where to take the momentum from.
In addition, the momentum sum rule is enforced with inclusion of only the inelastic photon for CT18$_{\rm QEDproton}$~\cite{Xie:2023qbn}, while CT18$_{\rm QEDfit}$ includes both inelastic and elastic components~\cite{Xie:2021equ}.
For MSHT we consider the corresponding QED sets for both the NNLO QCD~\cite{Bailey:2020ooq, Cridge:2021pxm} and aN3LO QCD PDFs~\cite{McGowan:2022nag, Cridge:2023ryv}. For the latter we take  {\tt MSHT20qed\_an3lo\_qcdfit} which is the QCD-only
counterpart to the QED PDFs (as slight updates were made relative to the original aN3LO QCD PDFs); this is denoted MSHT20$_{\rm QED\!\_QCD}$. In all cases in MSHT the LUXQED formula is applied at low scale and evolved in QCD$\otimes$QED evolution to the required scale, with the fit determining the PDFs and hence where the photon momentum is taken from~\cite{Harland-Lang:2019pla}. This is therefore most similar to the approach in CT18$_{\rm QEDfit}$.

\begin{table}[tb]
    \centering
\begin{tabular}{l|l|lll|l}
\toprule
PDF set & QCD order & $\langle x\Sigma\rangle$ & $\langle xg\rangle$ & $\langle x\gamma\rangle$ & Total \\
\midrule
\textbf{CT18} & NNLO & 0.5282 & 0.4718 & 0 & 1.0000 \\
$\mbox{CT18}_{\rm QEDproton}$ & NNLO & 0.5262 & 0.4708 & 0.00431 & 1.0013\textsuperscript{footnote }\tablefootnote{In CT18$_{\rm QEDproton}$~\cite{Xie:2021equ}, the momentum sum rule is enforced with the inclusion of inelastic photon, \emph{i.e.}, $\langle x(\Sigma+g+\gamma^{\rm ine})\rangle=1$, with the total slightly above one unit from the elastic component, $\langle x\gamma^{\rm el}\rangle=0.13\%$.} \\
$\mbox{CT18}_{\rm QEDfit}$ & NNLO & 0.5262 & 0.4694 & 0.00432 & 0.9999 \\
$\mbox{CT18}_{\rm LUX}$ & NNLO & 0.5282 & 0.4718 & 0.00437 & 1.0044\textsuperscript{footnote }\tablefootnote{In CT18$_{\rm LUX}$~\cite{Xie:2021equ}, direct application of Eq.~(\ref{eq:LUXQED}) produces a small total momentum excess.} \\
\hline
\textbf{MSHT20} & NNLO & 0.5305 & 0.4697 & 0 & 1.0001 \\
$\mbox{MSHT20}_{\rm QED}$& NNLO & 0.5279 & 0.4679 & 0.00436 & 1.0001 \\
\hline
$\mbox{\textbf{MSHT20}}_{\rm QED\!\_QCD}$ & aN3LO & 0.5379 & 0.4596 & 0 & 0.9975\textsuperscript{footnote }\tablefootnote{Note the slight differences from 1 for MSHT here reflect the LHAPDF extrapolation~\cite{Buckley:2014ana}.} \\
$\mbox{MSHT20}_{\rm QED}$ & aN3LO & 0.5358 & 0.4572 & 0.00441 & 0.9974\textsuperscript{footnote }\footnotemark[\value{footnote}] \\
\hline
\textbf{NNPDF3.1} & NNLO & 0.5294 & 0.4704 & 0 & 0.9998 \\
$\mbox{NNPDF3.1}_{\rm QED}$ & NNLO & 0.5271 & 0.4685 & 0.00435 & 0.9999 \\
\hline
$\textbf{NNPDF4.0}_{\rm QCD}$ & NNLO & 0.5341 & 0.4657 & 0 & 0.9998 \\
$\mbox{NNPDF4.0}_{\rm QED}$ & NNLO & 0.5330 & 0.4625 & 0.00433 & 0.9998 \\
\hline
\textbf{NNPDF4.0} & aN3LO & 0.5369 & 0.4629 & 0 & 0.9998 \\
$\mbox{NNPDF4.0}_{\rm QED}$ & aN3LO & 0.5362 & 0.4593 & 0.00433 & 0.9998 \\
\bottomrule
\end{tabular}
\caption{Momentum fractions for QCD-only (\textbf{bold}) as well as QED-enhanced PDF sets involving the photon density at $Q\! =\! M_H\! =\! 125$ GeV. See text for descriptions of the different PDF sets.}
\label{tab:MomFrac}
\end{table}
For NNPDF, we consider both NNPDF3.1~\cite{NNPDF:2017mvq} and 4.0~\cite{NNPDF:2021njg}, at NNLO\footnote{We note that for NNPDF4.0 the public QCD set differs slightly from that used as the baseline for the QED set~\cite{NNPDF:2024djq}, the corresponding PDF set used as the baseline is therefore taken from~\cite{NNPDF4QED_Website}.} in QCD for the former and also at aN3LO~\cite{NNPDF:2024nan} for the latter, and their corresponding QED-enhanced PDFs~\cite{Bertone:2017bme,NNPDF:2024djq,Barontini:2024dyb}. In NNPDF the photon PDF is determined iteratively by applying the LUXQED formula at the scale $Q=100~\textrm{GeV}$, and then running backward (forward) to a lower (higher) scale in the full QCD$\otimes$QED evolution, again with the fit determining the PDFs from where the photon momentum is taken. This is therefore most similar to the MSHT and CT18$_{\rm QEDfit}$ approaches, but with the difference that the LUXQED formula is now applied at a high scale. NNPDF, unlike CT or MSHT, does not include electroweak corrections for data in the baseline QCD PDFs; instead, they cut the data to avoid regions where electroweak corrections are relevant~\cite{NNPDF:2024djq}.

In order to demonstrate the impact of the inclusion of the photon on the momentum distribution of the partons through momentum conservation, we first consider the momenta carried by the quark-singlet, gluon, and photon partons in Table~\ref{tab:MomFrac} for this selection of NNLO and approximate N3LO (aN3LO) PDF sets ~\cite{Xie:2021equ,Bailey:2020ooq,Cridge:2021pxm,NNPDF:2017mvq,Bertone:2017bme,NNPDF:2021njg,NNPDF:2024djq,NNPDF:2024nan,Barontini:2024dyb}. In the rightmost column, we tally the corresponding values for the total momentum of the proton, for which results generally $=\!1$ within the working precision we quote; for those instances in which there are small, per-mille deviations from 1 in the total momentum, we provide footnotes and in-text discussion above. 
Overall, the net effect of the inclusion of the photon PDF is to reduce the gluon and quark-singlet momentum fractions of the PDFs very slightly (by $\approx\! 0.4\!-\!0.8\%$, and $\approx\! 0.1\!-\!0.4\%$ respectively) to generate a photon PDF with total momentum fraction consistently found by all groups to be $\langle x\gamma \rangle\! \approx\! 0.004$ ({\it i.e.}, $\approx$0.4\%).

\begin{figure}[htbp]
    \centering
    \includegraphics[width=7.0cm,height=4.8cm]{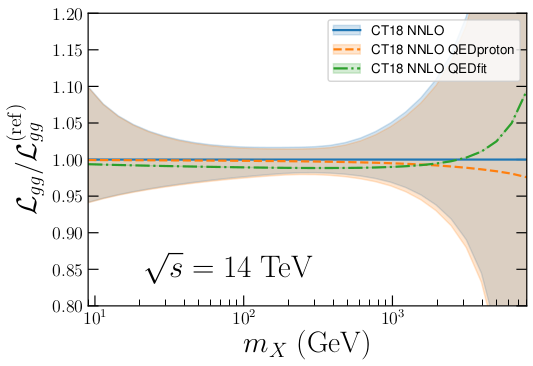}
    \includegraphics[width=7.0cm,height=4.8cm]{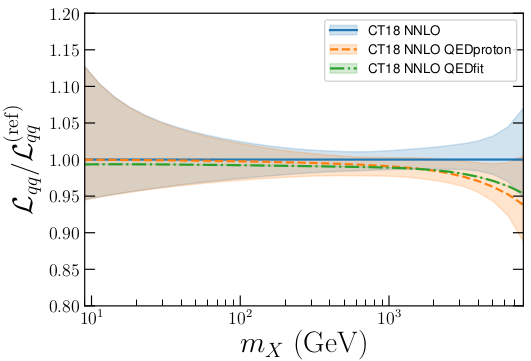} 
    \includegraphics[width=7cm,height=4.8cm]{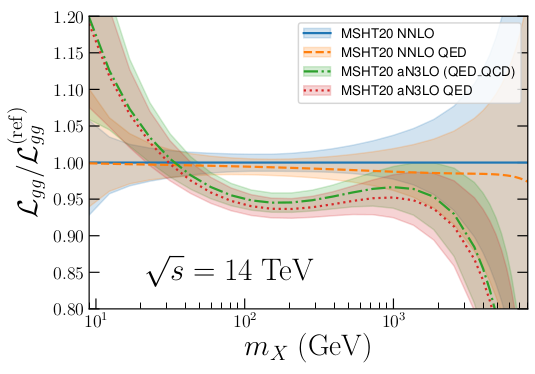}
    \includegraphics[width=7cm,height=4.8cm]{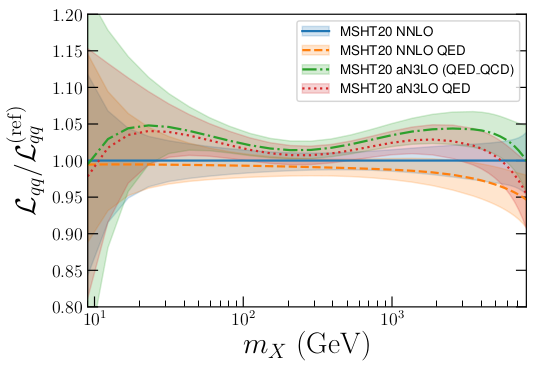} 
   
    \includegraphics[width=7cm,height=4.8cm]{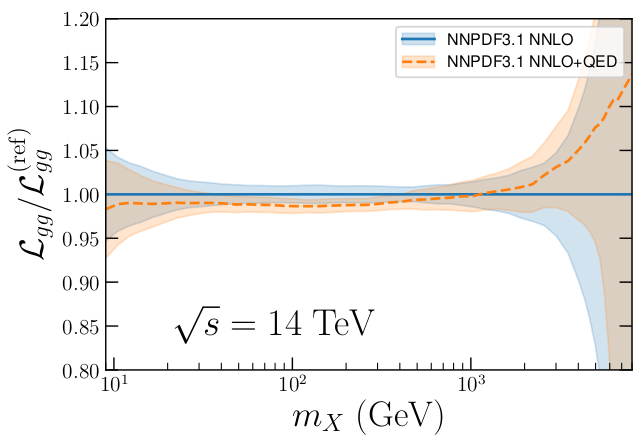}
    \includegraphics[width=7cm,height=4.8cm]{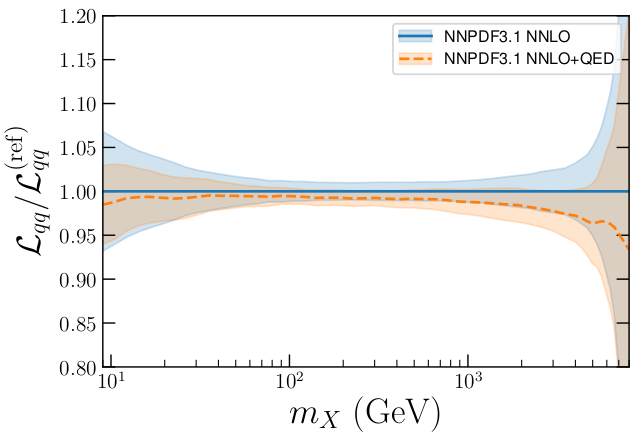} 
    \includegraphics[width=7cm,height=4.8cm]{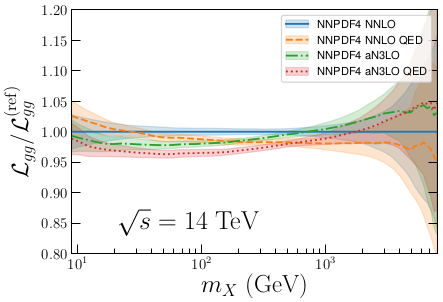}
    \includegraphics[width=7cm,height=4.8cm]{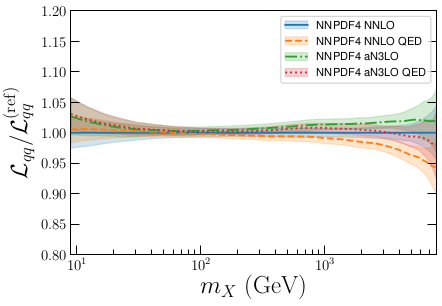}  \vspace{-5pt}
    \caption{The gluon-gluon (left column) and quark-quark (right column) parton luminosities for the CT18 (first row), MSHT20 (second row), NNPDF3.1 (third low) and NNPDF4.0 (fourth row) comparing QCD-only and corresponding QED-enhanced PDFs. The NNLO, NNLO+QED, and (where existing) aN3LO and aN3LO+QED PDFs are shown. In each panel, the plotted luminosities are normalized to a non-QED NNLO baseline (solid-blue curve) for each respective analysis effort. Corresponding plots for the quark-gluon and quark-antiquark combinations are shown in App.~\ref{app:lumis}.}
    \label{fig:PDF_lumis}
\end{figure}

The shifts in partonic momenta in Table~\ref{tab:MomFrac} as summarized above correspond to redistributions in the $x$ dependence of the quark-gluon PDFs which are then reflected in the $\sqrt{s}\!=\!14$ TeV parton luminosities in Fig.~\ref{fig:PDF_lumis}; there, all groups observe a reduction of $\lesssim\! 2\%$ in the gluon-gluon luminosity around $m_H$, with the largest effect observed in NNPDF4.0, both at NNLO and at aN3LO. This is also larger than seen in NNPDF3.1.
For NNPDF4.0, the results shown are for their NNLO/aN3LO PDF set without the estimated missing higher-order corrections (MHOU). The results for their corresponding PDF set with the MHOU are very similar and not shown.

The smallest effects are observed for the CT18$_{\rm QEDproton}$ set, in which the photon momentum is instead taken from the quark sea by hand. In comparison, the CT18$_{\rm QEDfit}$ set shows slightly smaller decreases than CT18$_{\rm QEDproton}$ for the quark luminosities at high invariant masses, {\it e.g.}, as seen in the quark-quark luminosity in the top right of Fig.~\ref{fig:PDF_lumis}, as the global fit favors a smaller pull from the QCD-only case in CT18 than that manually imposed in CT18$_{\rm QEDproton}$. 
In general, aside from this case, the impact on the quark luminosities of the inclusion of QED effects is usually slightly smaller than for the gluon luminosities. We also show the quark-gluon and quark-antiquark luminosities in Appendix~\ref{app:lumis}, the former largely follows the changes in the gluon, and the latter is more similar to the quark-quark case. 

\begin{figure}[htbp]
    \centering
    \includegraphics[width=7cm,height=5cm]{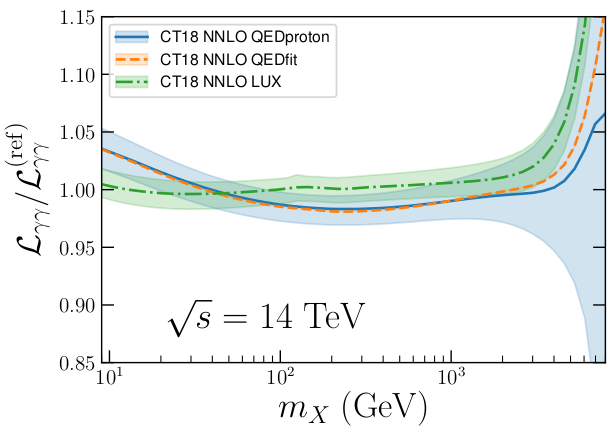}
    \includegraphics[width=7cm,height=5cm]{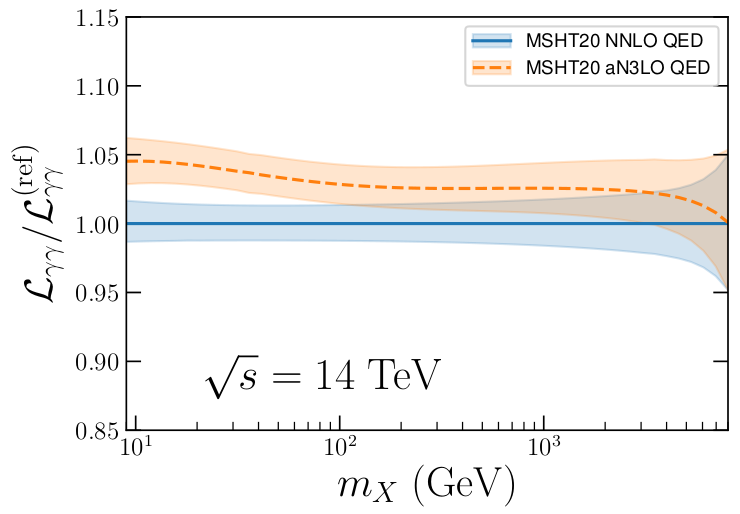}
    \includegraphics[width=7cm,height=5cm]{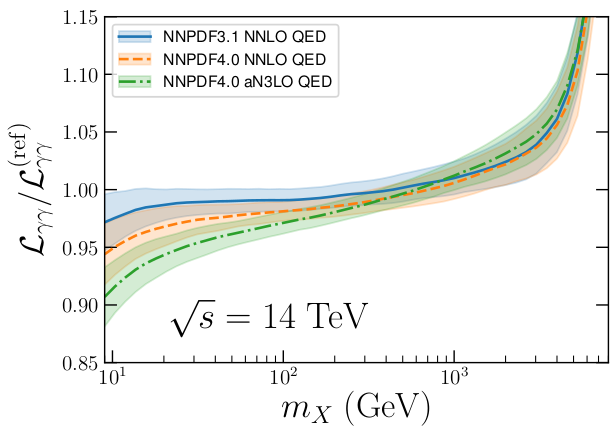} 
    \caption{The photon-photon luminosity for the NNLO+QED and aN3LO+QED PDF sets of MSHT20 (upper left), CT18 (upper right), and NNPDF3.1 and 4.0 (lower), shown throughout relative to a baseline of MSHT20 NNLO+QED, to which we normalize all curves as $\mathcal{L}^\mathrm{(ref)}_{\gamma\gamma}$.}
    \label{fig:PDF_photonlumis}
\end{figure}

We also compare the photon-photon luminosities of the different QED-enhanced PDF sets in Fig.~\ref{fig:PDF_photonlumis}, where we observe that the differences between PDF groups are largely at the level of a few-percent.
These differences typically follow differences in the charge-weighted singlet of the PDFs~\cite{Cridge:2023ryv}. When one considers that the overall magnitude of the photon PDF is very small, carrying total momentum of $\sim\! 0.4\%$ as indicated in Table~\ref{tab:MomFrac}, these remaining differences are therefore very small.

%
%

\section{QED-enhanced Higgs cross sections}
\label{sec:results}
Cross sections in the combined Higgs-electroweak
sector have potential sensitivity to the the precise treatment of the PDFs. This may occur either directly through the
presence and modeling of the proton's photon density and inclusion of NLO electroweak corrections to cross sections, or indirectly through the impact of these corrections and the QED-enhanced DGLAP evolution on the proton momentum distribution. For this purpose, first in Sec.~\ref{sec:total} we evaluate several
Higgs-production total cross sections using the \texttt{n3loxs}~\cite{Baglio:2022wzu} and \texttt{proVBFH-inclusive}~\cite{Dreyer:2018qbw} frameworks,
which allow calculations up to N3LO accuracy in $\alpha_s$.
For the Higgs-strahlung calculations, we also evaluate NLO electroweak effects using the \texttt{HAWK} package~\cite{Denner:2014cla}. As the higher-order QCD corrections
have been investigated in detail recently, we specifically concentrate on these
electroweak effects and their possible PDF dependence, particularly with respect
to recent analyses which have included the photon PDF in the LUX formalism.
As appropriate for Higgs-strahlung, the NLO electroweak theory we compute contains the complete range of corrections, which assume the form of virtual ({\it e.g.}, vertex and self-energy contributions) as well as real NLO effects as realized in the production of photons radiated to the final state. In addition, the inclusion of the full electroweak theory to NLO accuracy permits graphs consisting of photons in the initial-state of the hard scatter ({\it i.e.}, photon-initiated contributions), which depend directly on the photon PDF of the proton. 
We note that it is typically assumed that the higher-order electroweak corrections that
we explore effectively factorize from perturbative QCD contributions. Namely,
\begin{equation}
\frac{\mathrm{d} \sigma_{\text{QCD} \times \text{EW}}^{\text{best}}}{\mathrm{d} \mathcal{O}}
= \left[ 1 + \delta_{\text{EW}} (\mathcal{O}) \right] 
\frac{\mathrm{d} \sigma_{\text{QCD}}^{\text{best}}}{\mathrm{d} \mathcal{O}}
+ \frac{\mathrm{d} \sigma_{\gamma}}{\mathrm{d} \mathcal{O}}\ ,
\end{equation}
where $\sigma_{\gamma}$ above corresponds specifically to the photon-induced contributions
to the relevant observable(s). Furthermore,  it has been observed recently that the effect of adding QED evolution to NNLO or aN3LO QCD PDFs is quantitatively very similar~\cite{Cridge:2023ryv}, supporting the suggestion that the QED evolution effects factorize from the QCD order.

\begin{table}[tb]
    \centering
\begin{tabular}{l|l|l|l|l|l|l}
\toprule
PDF Set & QCD order & $\sigma_{gg \to H}$ & $\sigma_{\rm VBFH}$ & $\sigma_{Z H}$ & $\sigma_{W^+ H}$ & $\sigma_{W^- H}$ \\
\midrule
\textbf{CT18} & NNLO & 51.7 & 4.45 & 0.880 & 0.985 (0.904) & 0.626 (0.576) \\
$\mbox{CT18}_{\rm QEDproton}$ & NNLO & 51.6 & 4.43 & 0.876 & 0.980 (0.899) & 0.624 (0.574) \\
$\mbox{CT18}_{\rm QEDfit}$ & NNLO & 51.2 & 4.42 & 0.874 & 0.979 (0.898) & 0.622 (0.572) \\
$\mbox{CT18}_{\rm LUX}$ & NNLO & 51.8 & 4.45 & 0.880 & 0.985 (0.904) & 0.626 (0.576) \\
\hline
\textbf{MSHT20} & NNLO & 51.8 & 4.53 & 0.885 & 0.984 (0.903) & 0.624 (0.574) \\
$\mbox{MSHT20}_{\rm QED}$ & NNLO & 51.5 & 4.49 & 0.878 & 0.977 (0.896) & 0.619 (0.570) \\
\hline
$\mbox{\textbf{MSHT20}}_{\rm QED\!\_QCD}$ & aN3LO & 50.9 & 4.67 & 0.883 & 0.984 (0.903) & 0.624 (0.574) \\
$\mbox{MSHT20}_{\rm QED}$ & aN3LO & 50.3 & 4.64 & 0.877 & 0.975 (0.894) & 0.619 (0.570) \\
\hline
\textbf{NNPDF3.1} & NNLO & 52.8 & 4.49 & 0.900 & 1.010 (0.926) & 0.634 (0.583) \\
$\mbox{NNPDF3.1}_{\rm QED}$ & NNLO & 52.2 & 4.46 & 0.893 & 0.998 (0.915) & 0.630 (0.580) \\
\hline
$\textbf{NNPDF4.0}_{\rm QCD}$ & NNLO & 52.0 & 4.61 & 0.912 & 1.020 (0.936) & 0.642 (0.591) \\
$\mbox{NNPDF4.0}_{\rm QED}$ & NNLO & 51.3 & 4.60 & 0.909 & 1.010 (0.926) & 0.638 (0.587) \\
\hline
\textbf{NNPDF4.0} & aN3LO & 52.8 & 4.64 & 0.900 & 1.000 (0.917) & 0.632 (0.582) \\
$\mbox{NNPDF4.0}_{\rm QED}$ & aN3LO & 51.9 & 4.65 & 0.898 & 1.000 (0.917) & 0.630 (0.580) \\
\bottomrule
\end{tabular}
\caption{
Benchmark total cross sections [in $\mathrm{pb}$] for various processes of on-shell Higgs production at $\sqrt{s} = 14$ TeV, computed with indicated PDF sets and \texttt{n3loxs}, \texttt{proVBFH-inclusive}. For the latter $W^\pm H$ processes, we explicitly include the NLO electroweak-corrected cross section (in parentheses) as determined via \texttt{HAWK2.0}. We do not include in these cases a separate photon-initiated correction here, although this is discussed in Sec.~\ref{sec:HV} for the $W$-decayed $pp\to W^+ H$ process.}
\label{tab:xsecs}
\end{table}
%

%
%
\subsection{PDF-driven effects in total cross sections}
\label{sec:total}
We begin by collating the cross sections using a variety of QCD-only and QED-enhanced PDF sets for various Higgs total cross sections in Table~\ref{tab:xsecs}, these include gluon fusion Higgs production, vector boson fusion Higgs production and Higgs-strahlung. Among total cross sections, it is reasonable to expect both the VBF and Higgs-strahlung
processes to be more prominently affected by the inclusion of QED corrections. At the same
time, however, the inclusion of the photon PDF necessarily redistributes the total momentum
carried by the gluon and/or quark sea to the photon, resulting in reductions in cross sections from PDF effects. 

The largest of the Higgs production cross sections at the LHC is the gluon fusion, $gg \to H$, channel~\cite{LHCHiggsCrossSectionWorkingGroup:2016ypw,Spira:2016ztx,Jones:2023uzh,Karlberg:2024zxx}. One might suppose electroweak corrections to be fairly small for this process at the level of hard cross sections. 
However, in addition to these effects, gluon fusion may also be impacted indirectly through the subtle dependence of total cross sections on the shapes of PDFs;
careful examination can therefore reveal aspects of the relationship with global properties of the PDFs ---
particularly, the redistribution of momentum among active partons in QED-enhanced analyses. Indeed, it is clear from the third column of Table~\ref{tab:xsecs} that, while all groups consistently see a reduction in the $gg \rightarrow H$ cross section upon inclusion of QED effects, the magnitudes of this reduction vary.

\begin{figure}[htbp]
    \centering
    \includegraphics[width=0.75\textwidth]{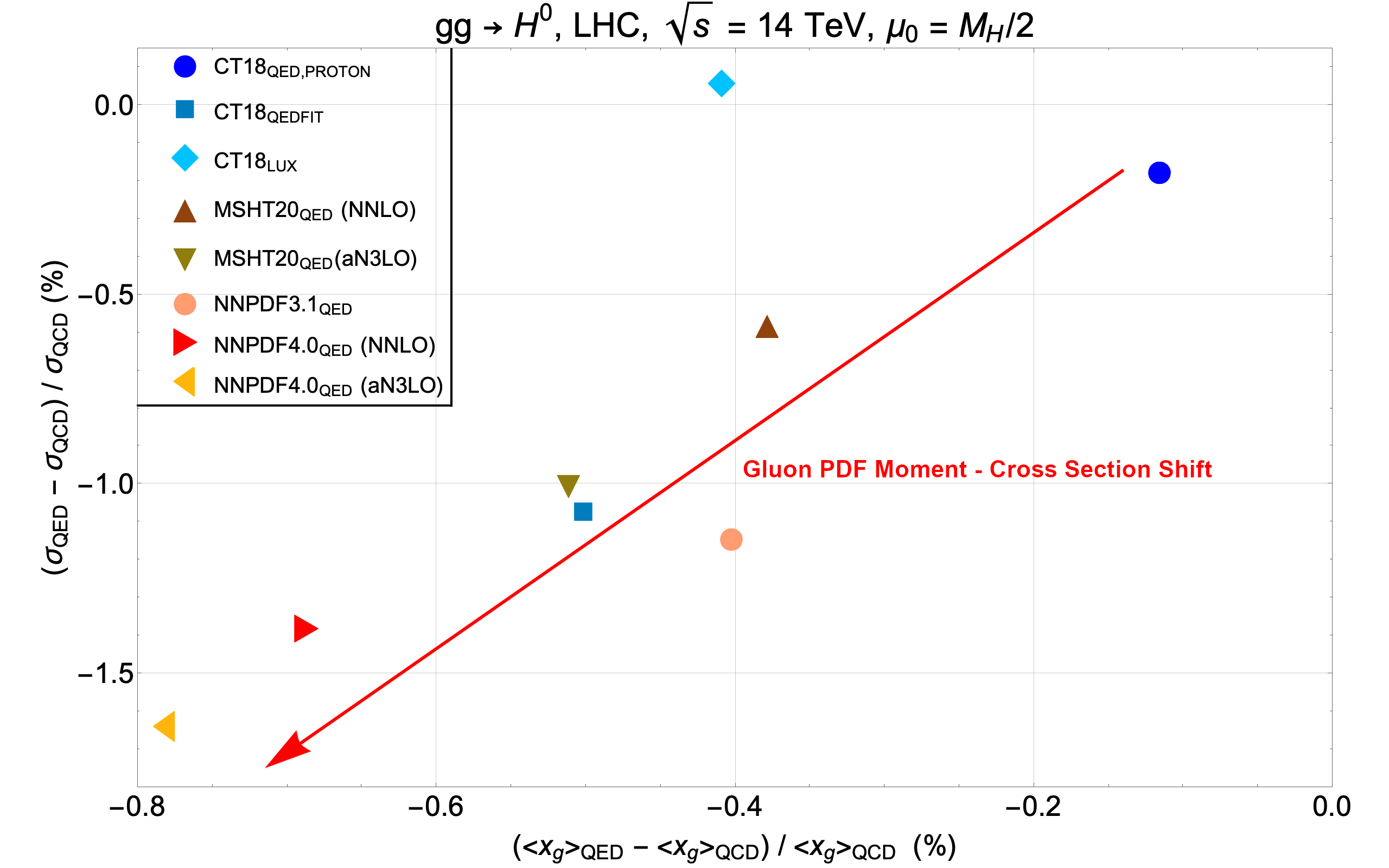}
    \caption{Correlations between the relative shifts in the total gg$\to$H cross section (vertical) and the gluon momentum fraction (horizontal); we compare predictions for QED-enhanced PDF sets against their QCD baseline counterparts.}
    \label{fig:ggH_tot}
\end{figure}

This can be visualized in Fig.~\ref{fig:ggH_tot}, where we plot the $gg \to H$ total cross section at 14 TeV with renormalization scale $\mu_0 = M_H / 2$.  We show on the vertical axis
the fractional deviation in the total cross section, when using the QED-improved PDFs, relative to their QCD-only baselines, relating this on the horizontal axis to the corresponding QED-induced
shifts in the momentum carried by the gluon distribution.

As described in Section~\ref{sec:PDFs}, including a photon PDF reduces the momentum carried by other partons in the proton, leading to a modest reshuffling of the PDFs and integrated moments of the QCD partons.  For the case shown here of gluon-fusion Higgs production, the dominant impact is a reduction in the gluon PDF (whose change in total momentum is shown on the abscissa), which in turn reduces the Higgs production cross section (as shown on the vertical axis). This is illustrated by the arrow in Fig.~\ref{fig:ggH_tot}, indicating the strong correlation of reduced gluon momentum with reduced $gg \rightarrow H$ cross sections. Moreover, the gradient of the line is $\sim2$, as the gluon fusion Higgs production cross section depends on the square of the gluon PDF at leading order.

In more detail, it is notable that the calculations based upon CT18 (in its ``QED fit'' version, which is most analogous to MSHT and NNPDF), MSHT20 and NNPDF3.1 at NNLO all sit at the center of the plot. They have comparable changes in their gluon momenta of $-0.3\%$ to $-0.5\%$ and corresponding shifts in the gluon fusion cross section of $-0.6\%$ to $-1.2\%$. On the other hand, as one might expect, CT18$_{\rm LUX}$ and CT18$_{\rm QEDproton}$ demonstrate little change in the gluon-fusion cross section, as there is only at most very limited momentum redistribution from the photon in these cases due to the choices made in those implementations.  NNPDF4.0 shows the opposite trend, with more significant decreases in the gluon momentum and, accordingly, the gluon fusion cross section, which may reflect differences in the NNPDF4.0 included datasets relative to NNPDF3.1~\cite{LesHouches2025}. We also note that the approximate N3LO PDF sets of MSHT and NNPDF4.0 both show slightly larger changes than their corresponding NNLO counterparts, albeit at a few-per-mille level. 

\begin{figure}[p]
    \centering
    \includegraphics[width=0.75\textwidth]{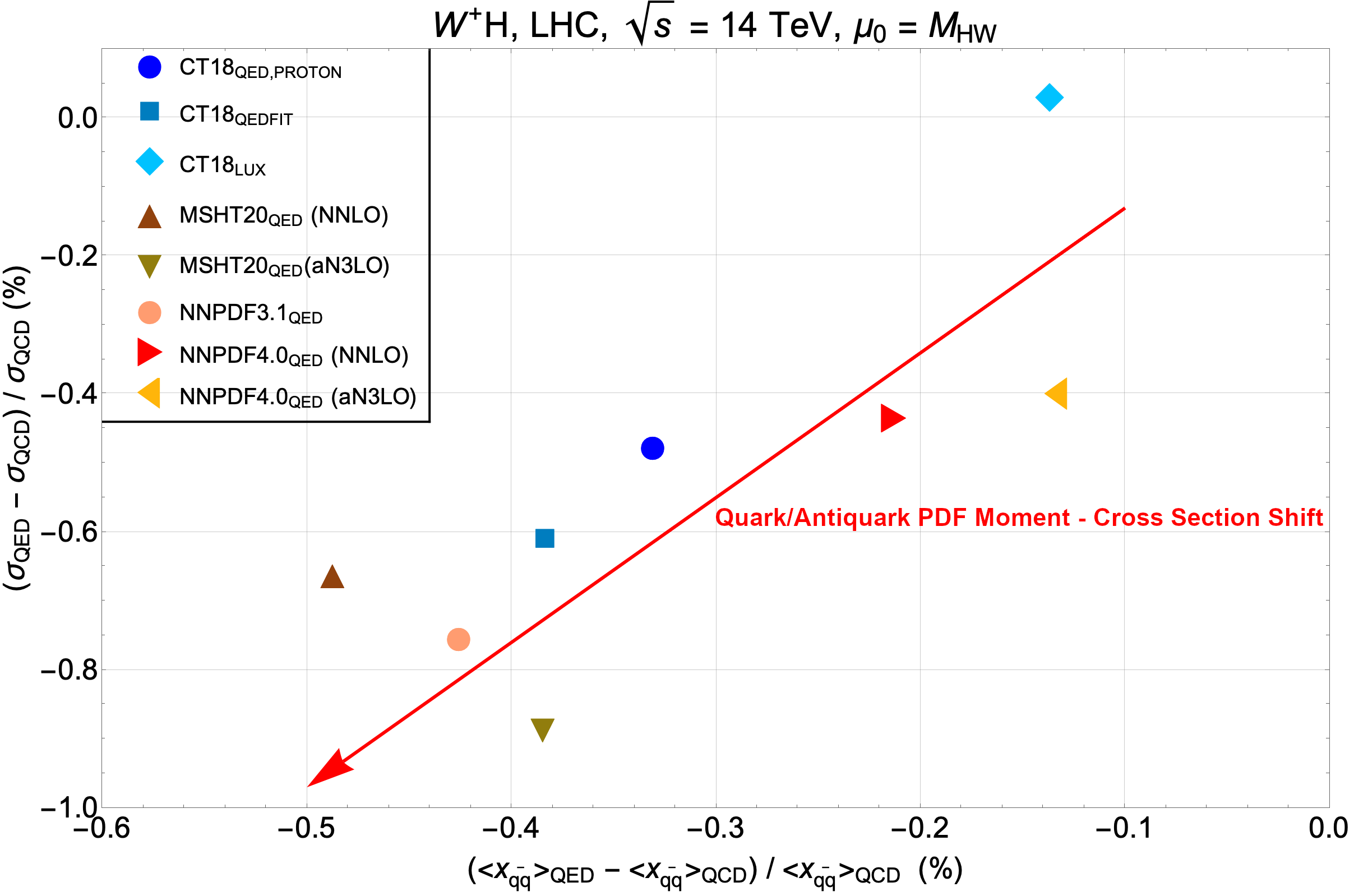}\\
    \vspace{1.0cm}\includegraphics[width=0.75\textwidth]{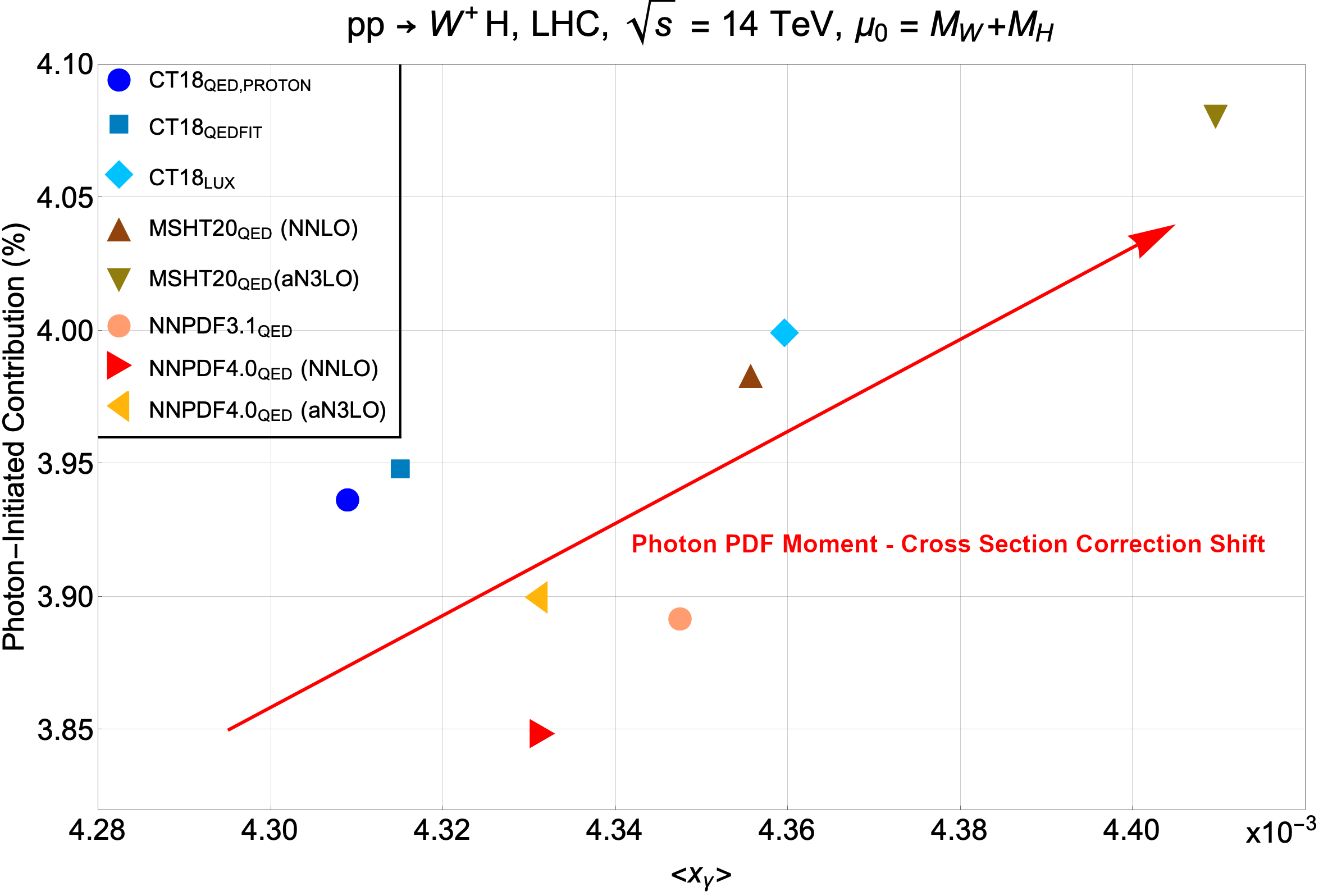} 
    \caption{Same as Fig.~\ref{fig:ggH_tot}, for (upper) correlations of relative shifts in the total on-shell $W^+H$ cross section without NLO electroweak corrections (vertical) and quark + antiquark momentum fractions (horizontal) and (lower) correlations of photon-induced contributions (vertical) to the total $(W^{+} \to \ell^+\nu) H$ cross section and the total photon momentum fractions (horizontal). The cross sections are computed for the LHC at $\sqrt{s}=14$ TeV with a dynamical scale equal to the $WH$ invariant mass in the upper panel and fixed to $M_W\! +\! M_H$ in the lower one.}
    \label{fig:WpH_tot}
\end{figure}

Turning next to consider Higgs-strahlung $pp\to WH$, we expect another pattern of electroweak and QED effects. Firstly, to illustrate purely the indirect effects of the QED evolution of the PDFs, the upper Fig.~\ref{fig:WpH_tot} shows the fractional change in the total $W^+H$ cross section on the vertical axis, computed by \texttt{n3loxs} with renormalization scale $\mu_0 = M_{WH}$, the invariant mass of the $WH$ system, now plotted against the change in the quark-antiquark-singlet momentum fraction on the abscissa. Again, we observe that the PDF sets of CT18 (QED fit), MSHT20 and NNPDF3.1 at NNLO demonstrate similar effects, with changes of $-0.4\%$ to $-0.5\%$ in their quark-antiquark momenta and corresponding impacts of $-0.6\%$ to $-0.8\%$ on the $W^+H$ total cross section. CT18$_{\rm LUX}$ again shows very little changes due to the fact that there is negligible redistribution of the photon momentum here (the photon is simply added on top of the QCD partons). On the other hand, CT18$_{\rm QEDproton}$, where the photon momentum is taken by hand from the quark sea, shows a more notable reduction in the quark-antiquark momentum, and hence also in the Higgs-strahlung cross section. NNPDF4.0 now shows smaller drops in the quark-antiquark momenta compared to the other PDF ensembles, perhaps related to their different selection of data compared to NNPDF3.1~\cite{LesHouches2025} and the resulting larger impact it sees in reducing the gluon momenta.

However, there are also significant NLO electroweak corrections to Higgs-strahlung, particularly at large transverse momenta.
We therefore supplement the on-shell production cross sections discussed so far with comparisons of respective cross
sections that also include decay effects, namely, in the $pp\to (W^+ \to \ell^+ \nu_\ell)\,H$ channel, for which we
assess electroweak corrections. We compute these cross sections at NLO electroweak accuracy 
using \texttt{HAWK2.0}~\cite{Denner:2014cla} code at the scale $\mu_0 = M_{W}\!+\!M_{H}$.
NLO electroweak considerations may potentially change the PDF effects on the cross section, as both photon-initiated contributions and NLO electroweak corrections contribute to the cross section, and we thus see shifts for all cross sections relative to the upper Fig.~\ref{fig:WpH_tot}. We therefore present in the lower panel of Fig.~\ref{fig:WpH_tot} a plot showing on the abscissa the total photon momentum of the different PDF sets, against the photon-initiated contribution to the cross section on the vertical axis.
In general, the impact of the photon-initiated contribution is at the level of 4\% of the total cross sections in the leptonic decay channel, but with a high level of consistency across predictions based on different QED PDFs.
As such, we observe a significant correlation between the photon-initiated contribution and total photon momentum, but with PDF-driven differences among the photon-initiated contributions of at most few per mille and less for the NNLO versions of CT18 (QED fit), MSHT20 and NNPDF3.1. Due to the minimal PDF dependence in this photon-initiated piece, the spread in total cross sections shown in Fig.~\ref{fig:WpH_tot} (upper) is predominantly governed by the redistribution of parton momenta entering the calculation before NLO electroweak corrections (including $\gamma$-initiated contributions) are introduced.

\begin{figure}[htbp]
    \centering
    \includegraphics[width=0.75\textwidth]{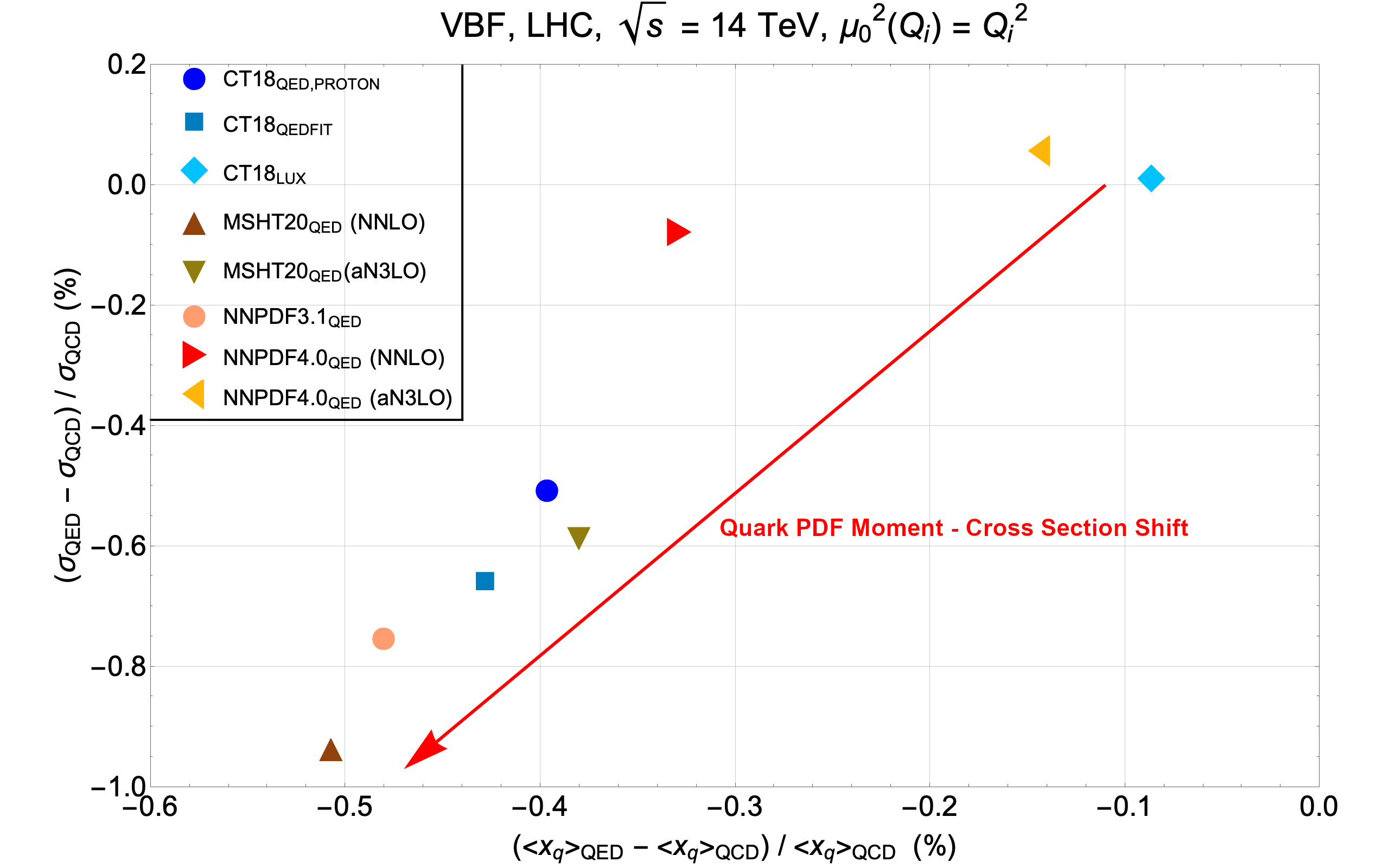}
    \caption{Correlation of the relative shifts in the total vector boson-fusion Higgs (VBFH) cross section (vertical) and quark momentum fractions (horizontal), comparing predictions from the various QED-enhanced PDF sets against their QCD baseline counterparts.}
    \label{fig:VBF_tot}
\end{figure}

Finally, in Fig.~\ref{fig:VBF_tot} we consider vector-boson fusion production of the Higgs boson (VBFH) with the dynamical scale, $\mu_0^2 = Q_i^2$ ($i=1,2$), {\it i.e.}, the respective virtualities of the exchange bosons in the structure function approach to VBFH~\cite{Dreyer:2018qbw}. As VBFH has quark-quark initiated channels, we plot the QED PDF-induced shifts in the total cross section on the vertical axis now against the change in the total quark PDF momentum on the horizontal axis. Once more, the reduction in momentum in the quarks results in a corresponding reduction of the total cross section upon the addition of QED effects. It is again noteworthy that the CT18 (QED fit), MSHT20 and NNPDF3.1 NNLO PDFs show very similar behavior with all having changes of the quark momentum of within less than one per mille of $-0.45\%$ and corresponding shifts in the total vector boson fusion production cross sections of $-0.6\%$ to $-1.0\%$. As before, CT18$_{\rm LUX}$ shows very little change as expected, while CT18$_{\rm QEDproton}$ shows reductions in the quark momentum and cross section similar to, though slightly smaller than, CT18$_{\rm QEDfit}$. NNPDF4.0 also again shows smaller differences, possibly for similar reasons as noted above for Higgs-strahlung and the quark-antiquark momenta. The approximate N3LO PDFs show slightly reduced differences from the addition of QED effects relative to the NNLO PDFs, consistent with the larger changes seen in the gluon fusion case and the need for total proton momentum conservation.

%
%
\subsection{Differential predictions: $W$-associated Higgs production}
\label{sec:HV}

\begin{figure}[htbp]
    \centering
    \includegraphics[width=0.495\textwidth]{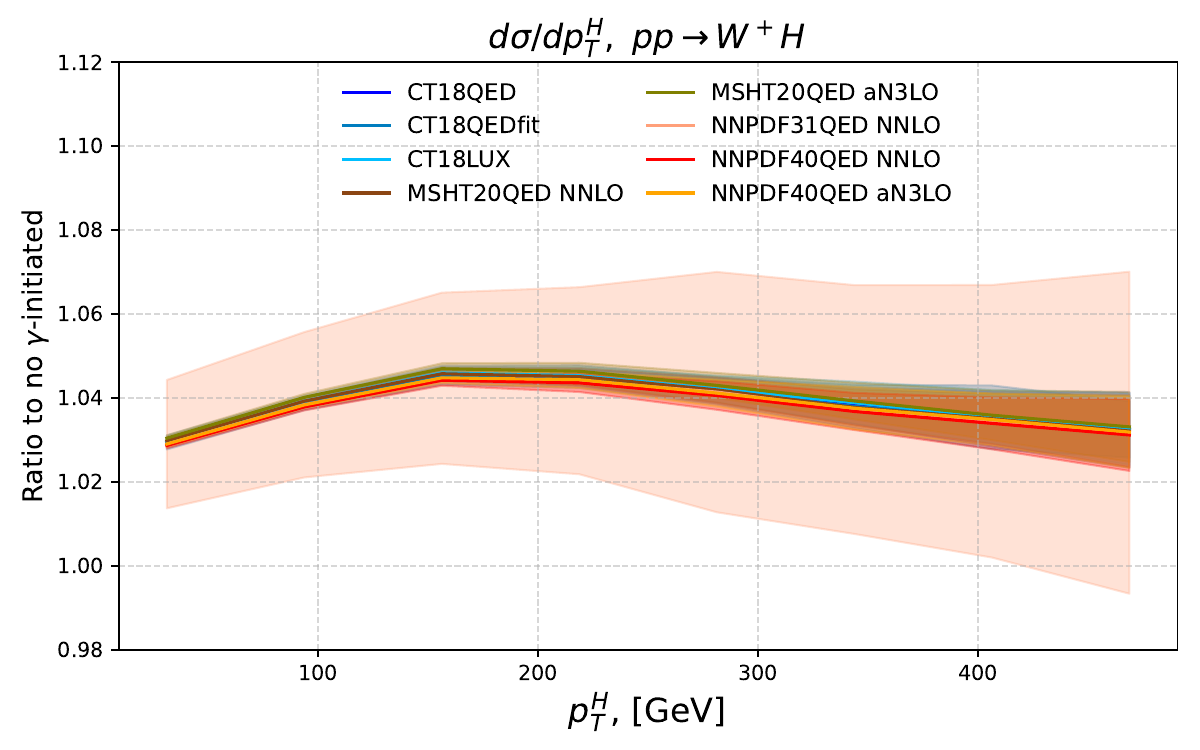}
    \includegraphics[width=0.495\textwidth]{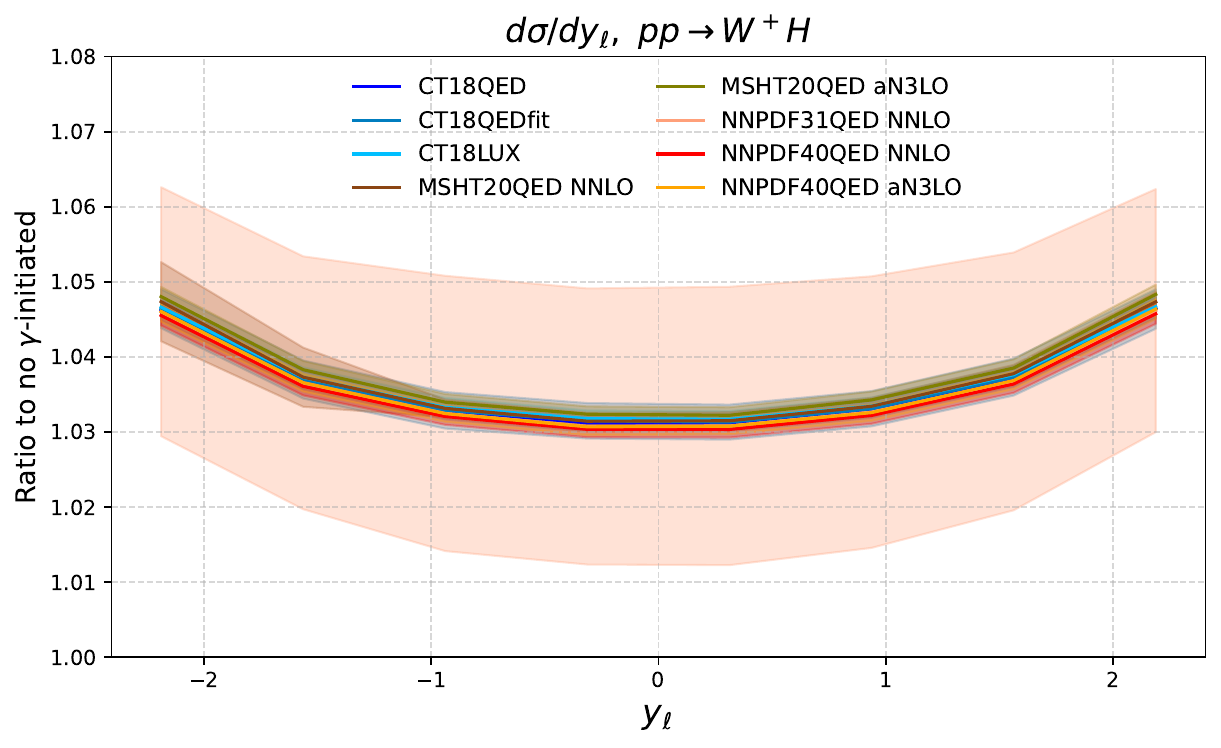}
    \caption{
	    The ratio of the full electroweak NLO-corrected $pp \to W^+H$ cross section, differential in $p^H_T$ (left) and
	    the charged lepton rapidity, $y_{\ell^+}$ (right), relative to the
        corresponding calculation {\it without} the photon-initiated contributions. The
        high degree of similarity illustrates the minimal dependence of the
        photon-initiated contribution to the specific QED-enhanced PDF, whether fitted at
        NNLO in $\alpha_s$ or approximate N3LO.
}
    \label{fig:higgs-diff-5}
\end{figure}

In addition to the total cross sections explored in Sec.~\ref{sec:total} above, it is also worthwhile to
investigate the electroweak corrections to singly differential Higgs-production cross sections.
Quantifying the possible effects in differential distributions as driven by variations in the
QED-enhanced PDFs complements the analogous study of the total cross section, given that the total cross sections
are more immediately correlated with the integrated PDF moments shown in
Fig.~\ref{fig:ggH_tot}--\ref{fig:VBF_tot}.
In contrast, the kinematical dependence in differential distributions supplies information
with a direct relation to the underlying $x$ dependence of the QED-enhanced PDFs upon which
the cross-section calculations are based.
This observation can be inferred from the Born-level kinematical matching(s) associated with
the corresponding collider predictions.
Generically, the cross section for production of a final-state electroweak particle, $B$, at the LHC, $p_1 + p_2 \to B + X$, might be matched to
\begin{equation}
x_{1,2} \sim {M_{B} \over \sqrt{s}}\, \exp\{\pm y_{B}\}\ , \ \ \ \ \ Q \sim M_{B}\ ,
\label{eq:kinematics}
\end{equation}
with this matching subject to higher-order corrections motivating the inclusion of greater theoretical
working accuracy in both $\alpha_s$ and the electroweak interaction strength. For instance,
from Eq.~(\ref{eq:kinematics}) we conclude that rapidity-dependent cross sections,
$d\sigma / dy_{B}$, involve an $x$ dependence spanning well beyond the nominally expected
low-$x$, with forward-boosted electroweak events probing the high-$x$ behavior of the PDFs,
in addition to possible BSM contributions.
Complementary distributions like $p_T$ spectra can similarly furnish PDF-sensitive information, with the large-$p_T$
tails measured in two-body final states probing $x\! \sim\! 2p_T / \sqrt{s}$ within the central-rapidity region.
\begin{figure}[htbp]
    \centering
    \includegraphics[width=0.495\textwidth]{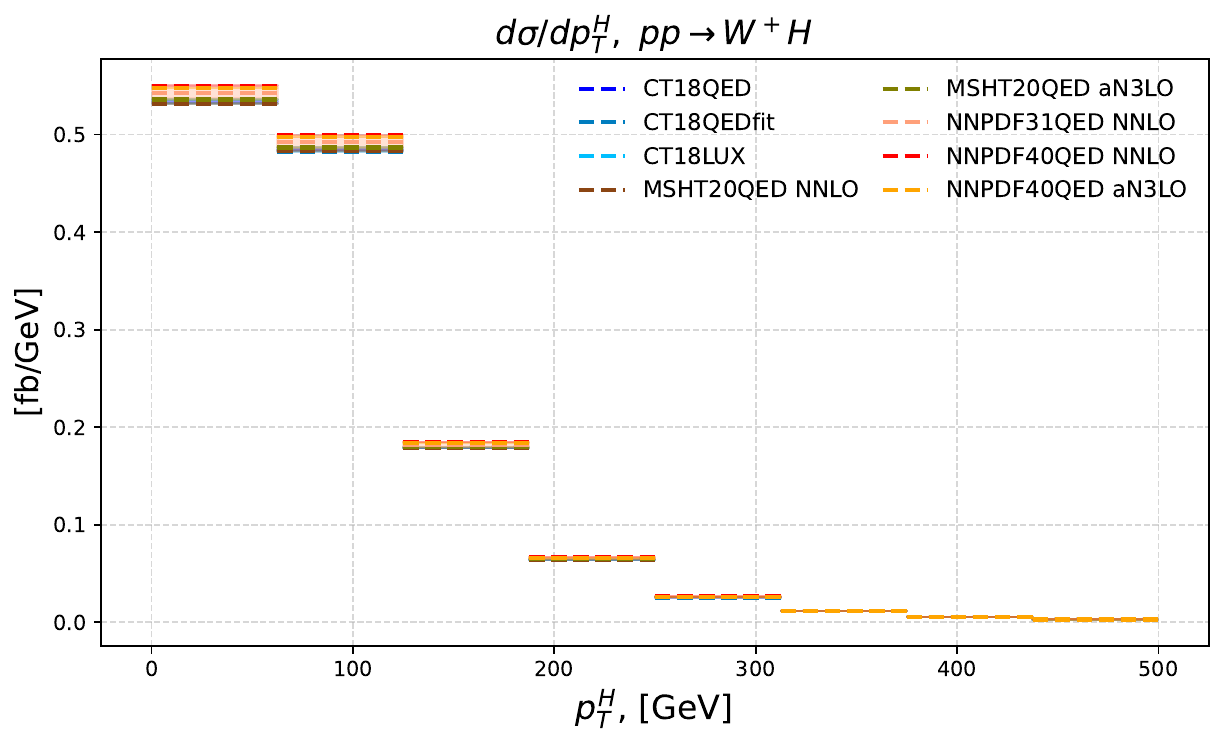}
    \includegraphics[width=0.495\textwidth]{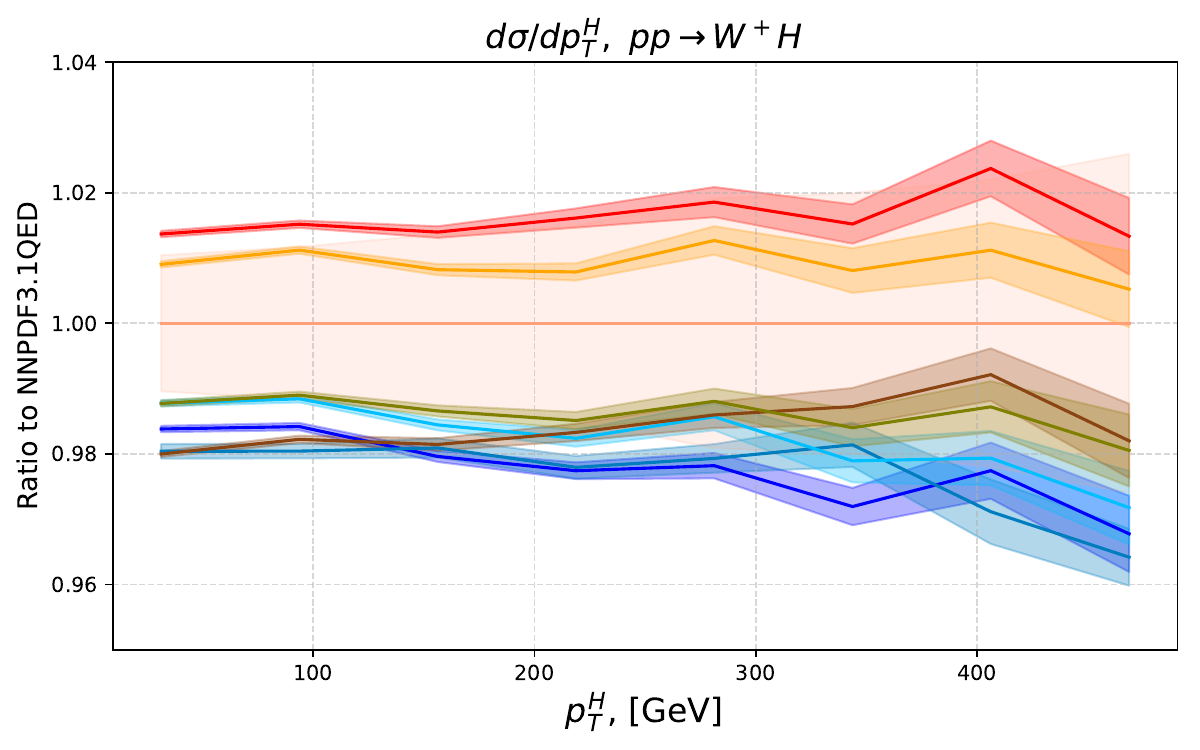}
    \caption{
	    The $pp \to W^+H$ cross section, differential in
	    the Higgs transverse momentum, $p^H_T$, illustrating
	    the QED-enhanced PDF dependence based on global fits
	    which included the photon PDF and (N)NNLO corrections in
	    $\alpha_s$. The left panel shows the absolute $p^H_T$ spectrum, while the right normalizes each prediction to the NNPDF3.1QED result, for which we also plot the PDF uncertainty. Here, and below, we assume the nominal scale-choice for associated production, $\mu_0\! =\! M_W\! +\! M_H$.
}        
    \label{fig:higgs-diff-3}
\end{figure}
Within this context, we select a particular Higgs-production process from those explored
above and examine the interplay of NLO electroweak corrections with the chosen QED-enhanced
PDFs.
Doing so allows us to quantify the size of NLO electroweak effects for a typical Higgs-production
process, particularly in light of the recent battery of new photon-PDF determinations summarized in Sec~\ref{sec:PDFs}.
As a representative example, we take the $W^+$-associated Higgs process, $pp \to W^+ H$, and compute
several quantities derived from singly differential cross sections in
Figs.~\ref{fig:higgs-diff-3}--\ref{fig:higgs-diff-5}.
We highlight the behavior in $W^+$ Higgs-strahlung again given the significant contribution of photon-initiated graphs to this process at NLO electroweak accuracy --- a point we observed for the total cross sections.
We are further motivated by the fact that the tails of Higgs-strahlung distributions have been underscored as potentially fertile search grounds for various BSM scenarios; for instance, anomalous trilinear Higgs couplings can potentially shift transverse-momentum
spectra of the Higgs in $VH$ processes, both at relatively low ($p^H_T \lesssim 100$ GeV)
and high momenta~\cite{Maltoni:2017ims}.
We note that we assume nominal (fixed) scale choices across all ($W$-decayed) $W^+H$ calculations shown below; namely,
$\mu_0\! =\! M_W\! +\! M_H$. All (integrated) cross sections are given in [$\mathrm{fb}]$, such that, {\it e.g.}, $p^H_T$ spectra are given in [$\mathrm{fb}/\mathrm{GeV}$], {\it etc}.
\begin{figure}[htbp]
    \centering
    \includegraphics[width=0.495\textwidth]{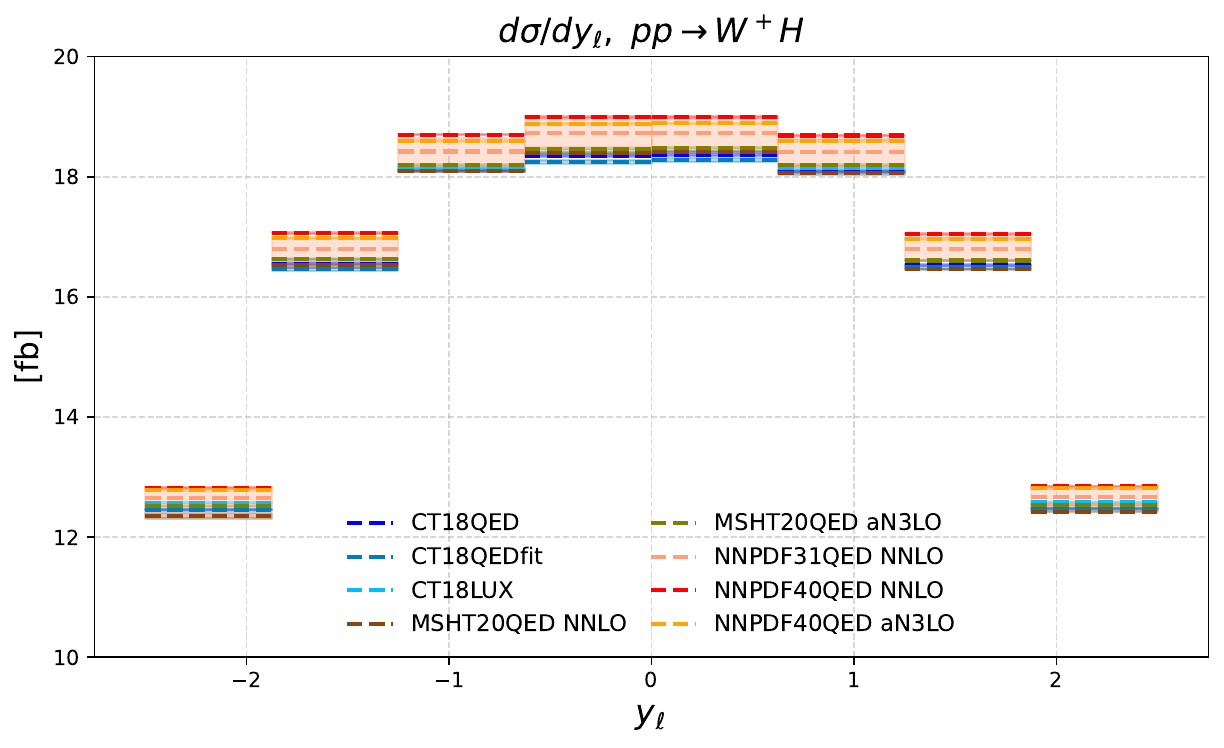}
    \includegraphics[width=0.495\textwidth]{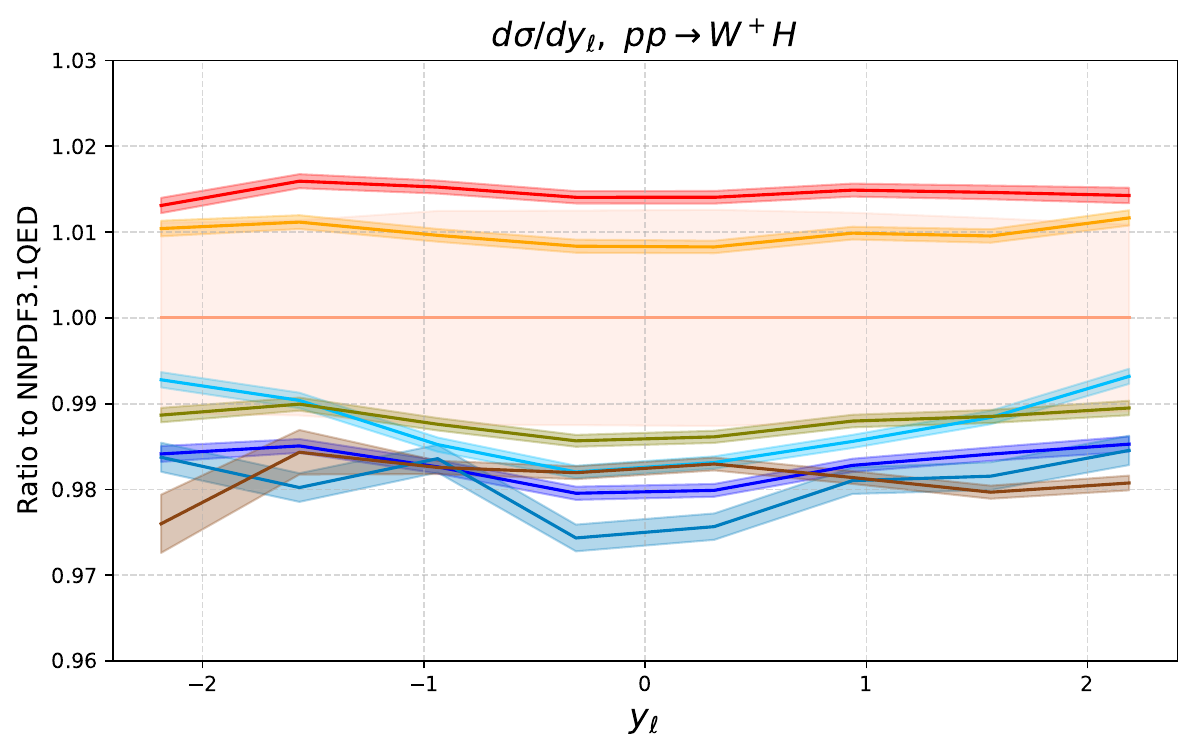}
    \caption{
	    The $pp \to W^+H$ cross section, differential in
	    the charged lepton rapidity, $y_{\ell^+}$ now illustrating
	    the QED-enhanced PDF dependence as in Fig.~\ref{fig:higgs-diff-3}; as before, the absolute differential cross section is shown (left) as well as the ratio of each calculation relative to that based upon NNPDF3.1QED (right).
}
    \label{fig:higgs-diff-4}
\end{figure}

To explore the dependence of differential $W^+$ Higgs-strahlung cross sections, we make consistent use of the \texttt{HAWK2.0} code, leveraging the full perturbative accuracy available internally within that framework --- namely, NLO in the electroweak theory and QCD. (This is in contrast to the numerical results given in Table~\ref{tab:xsecs}, for which we simply applied a relative NLO electroweak $K$-factor extracted from \texttt{HAWK2.0} to the \texttt{n3loxs} total cross sections.) By default for the results in this section, we therefore evaluate the full NLO electroweak calculation, {\it i.e.}, the contributions from virtual electroweak loop corrections as well as the photon-initiated pieces entering at NLO accuracy.
As such, we compute the $W$-decayed differential predictions which are the \texttt{HAWK2.0} default output, $pp \to (W^+\! \to\! \ell^+\nu)H$. We stress that these $W$-decayed predictions can ultimately be related unambiguously to the total $pp\to W^+H$ cross sections of Table~\ref{tab:xsecs} given knowledge of the NLO leptonic branching fractions of the $W^+$, assuming photon-initiated contributions are neglected~\cite{LHCHiggsCrossSectionWorkingGroup:2016ypw}. We therefore take these default settings and compute Higgs-$p_T$ spectra and $y_\ell$-rapidity distributions from the decayed $W^+$; for these predictions, we assume a default $\sqrt{s}=14$ TeV and take cuts and parameter selections as in Ref.~\cite{Obul:2018psx}, the results of which we are able to reproduce exactly. In general, we then highlight the behavior and QED-enhanced PDF dependence of the differential cross sections at NLO electroweak accuracy.

In Fig.~\ref{fig:higgs-diff-3}, we start by plotting the Higgs-$p^H_T$ spectrum, $d\sigma/dp^H_T$. In the left panel, we show the absolute spectrum for $p^H_T\! <\! 500$ GeV, computing the singly differential cross section for each of the eight variations among the QED-enhanced PDF sets explored in earlier sections. As in subsequent plots, we deploy a coarse binning to minimize Monte Carlo uncertainties while illustrating the qualitative dependence on kinematical variables. Across the plotted family of predictions based on different PDFs, Fig.~\ref{fig:higgs-diff-3} (left) illustrates a strong level of agreement in the rapid falloff of the differential cross section, which decreases by more than an order-of-magnitude from its peak by the latter $p^H_T$ bins at $p^H_T\! \ge\! 250$ GeV.

The robust agreement in the predictions among the various QED PDFs is still more apparent in the right panel of Fig.~\ref{fig:higgs-diff-3}, wherein we plot the ratio of each $p^H_T$ spectrum relative to the NNLO prediction based on NNPDF3.1QED (pink). The plotted bands for each prediction represent a Monte Carlo uncertainty, which becomes relatively sizable in the high-$p^H_T$ bins for which the cross section is small. Uniquely for the NNPDF3.1QED baseline, we plot its PDF uncertainty, which is approximately $\gtrsim\! 1\%$ at smaller $p^H_T$ before growing to $\sim\!2\%$ at $p^H_T\! \gtrsim 400$ GeV.
Notably, the PDF uncertainty in the $p^H_T$-dependent cross section can be identified with that of the total ($W^+$-decayed) cross section, for which we obtain
$\delta^{\mathrm (PDF)} \sigma_{W^+H}\! =\! 1.16\%$, comparable to the corresponding uncertainty in the low-$p^H_T$ region which dominates the integrated spectrum.
Regarding the PDF-driven spread in predicted $p^H_T$ spectra, we find the eight QED-enhanced sets approximately bracket the 1$\sigma$ PDF uncertainty of the NNPDF3.1QED prediction, with the NNLO and aN3LO NNPDF4.0QED predictions yielding the largest and next-to-largest predictions, respectively. In contrast, the three CT (CT18QED, QEDfit, and LUX) and two MSHT20QED (NNLO and aN3LO) predictions all have comparable magnitude and shape in $p^H_T$, being approximately $\sim\!2\%$ smaller than NNPDF3.1QED. Overall, we observe that the most up-to-date QED-enhanced PDFs induce a $\sim\!3\!-\!4\%$ spread in the $W^+$-associated Higgs production $p^H_T$ spectrum, with good concordance up to PDF uncertainties. We also underscore that this relative agreement has emerged in the wake of adopting the LUX formalism for the photon PDF.

In Fig.~\ref{fig:higgs-diff-4}, we show plots analogous to those of Fig.~\ref{fig:higgs-diff-3}, but now for the rapidity of the decay-lepton, $W^+ \to \ell^+\nu$. As before, we show the absolute rapidity distribution in the left panel, $d\sigma/dy_\ell$, over the same ensemble of QED-enhanced PDF sets. Correspondingly, the right panel of Fig.~\ref{fig:higgs-diff-4} illustrates the relative variation in these predictions with respect to the NNPDF3.1QED baseline, again giving the PDF uncertainty of the NNPDF3.1QED prediction for comparison against the spread in PDF predictions. We note the same qualitative behavior as in the $p^H_T$ spectra, with a fairly weak kinematical dependence and clear ordering in the magnitude of the predicted cross sections, for which the NNPDF4.0QED results are $\sim\!1\!-\!2\%$ larger than the NNPDF3.1QED baseline whereas the predictions based on the QED-enhanced CT and MSHT sets are comparatively smaller, by $\sim\!1\!-\!2\%$. As before, these predictions lie at the periphery of the $\sim\!1\sigma$ NNPDF3.1QED PDF uncertainty, which is largest ($\gtrsim\!1\%$) in the central-rapidity region before tapering slightly to $\sim\!1\%$ for $|y_\ell| \ge 2$.
The results plotted in Fig.~\ref{fig:higgs-diff-3} and~\ref{fig:higgs-diff-4} contain full NLO electroweak corrections, including photon-initiated contributions. Given the role of photon-initiated sub-processes in Higgs-strahlung, it is also valuable to investigate their relative contribution to the full NLO differential calculation. We show this in Fig.~\ref{fig:higgs-diff-5}, which plots the ratio of the full NLO result to that in which photon-initiated processes have been excluded. The resulting fraction quantifies the share of photon-initiated contributions to the full NLO-corrected calculation. As counterparts to Figs.~\ref{fig:higgs-diff-3} and~\ref{fig:higgs-diff-4}, the left and right panels of Fig.~\ref{fig:higgs-diff-5} provide this ratio for the $p^H_T$ spectrum and $y_\ell$ distribution, respectively. In both cases, we find the photon-initiated contribution represents $\sim\!3\!-\!4\%$ of the full NLO result, with evidence of some mild kinematical dependence; for instance, the $\gamma$-contribution to the $p^H_T$ spectrum peaks at $\gtrsim\!4\%$ near $p^H_T\! \approx\! 150$ GeV, while the corresponding contributions to the $y_\ell$ distributions are greatest in the more forward or backward rapidities, potentially exceeding $4.5\%$.
Strikingly, although we observe a significant impact from photon-initiated contributions in Fig.~\ref{fig:higgs-diff-5}, we note that the {\it variation} in central predictions for this impact with different assumed PDFs is very small, essentially spanning only several per mille; this behavior is analogous to the weak PDF dependence noted for the corresponding total cross sections shown in Fig.~\ref{fig:WpH_tot} (lower). This PDF independence of the photon-initiated contribution can be seen in both panels of Fig.~\ref{fig:higgs-diff-5}, with the spread in central predictions significantly overshadowed by the $\sim\!2\%$ nominal PDF uncertainty of the cross section ratio itself, which we representatively show for NNPDF3.1QED as before.
From these results, we conclude that the post-LUX QED PDF dependence of NLO corrections can be very mild --- generally at a per-mille level; this is true for both the virtual NLO electroweak corrections as well as the photon-initiated contributions plotted above, both of which we have quantified specifically for $pp\! \to\!W^+H$.
We emphasize that this observed PDF independence holds particularly for the NLO electroweak corrections themselves in the case of Higgs-strahlung, and reinforces precision in the perturbative evaluation of such electroweak effects. At the same time, however, it should be kept in mind that, with or without NLO electroweak corrections, {\it the PDF dependence of full cross sections remains significant}, as seen in the $\sim\!3\!-\!4\%$ spreads in the right panels of Fig.~\ref{fig:higgs-diff-3}--\ref{fig:higgs-diff-4} or the outer pink band (the NNPDF3.1QED PDF uncertainty) shown in both panels of Fig.~\ref{fig:higgs-diff-5}.
%

%
%
\subsection{PDF4LHC: observations and recommendations}
\label{sec:PDF4LHC}

Based on the results in the previous sections, we now summarize an estimate of QED corrections for key LHC cross sections based on the combined PDF4LHC21 NNLO PDF ensemble \cite{PDF4LHCWorkingGroup:2022cjn}. Table~\ref{tab:xsecsPDF4LHC} lists, first, the same cross sections as in Table~\ref{tab:xsecs}, now computed using the central PDF set of the 40-member PDF4LHC21 NNLO PDF ensemble. 
These QCD-only cross sections are very close\footnote{The reason they are not exactly the numerical average is that the CT18 and NNPDF3.1 contributions to PDF4LHC21 were slightly updated relative to the public ones to ensure consistent heavy quark masses. In the latter case, the data sets were also updated~\cite{PDF4LHCWorkingGroup:2022cjn}.} numerically to the averages of the respective central cross sections for CT18 NNLO, MSHT20 NNLO, and NNPDF3.1 NNLO PDFs in Table~\ref{tab:xsecs}, since 
the central PDF4LHC21 PDFs are given by the unweighted averages of CT18, MSHT20, and NNPDF3.1 ones entering the PDF4LHC21 combination, {\it i.e.},
\begin{equation}
    f_{a}(x,Q_0)_\textrm{PDF4LHC21} =\frac{1}{3}\left( f_{a}(x,Q_0)_\textrm{CT18} + f_{a}(x,Q_0)_\textrm{MSHT20} + f_{a}(x,Q_0)_\textrm{NNPDF3.1} \right). 
    \label{PDF4LHC21central}
\end{equation}
We can estimate the QED PDF corrections to the QCD-only PDF4LHC21 cross sections according to a similar formula, 
\begin{equation}
    \delta^\textrm{QED}_\textrm{PDF4LHC21} =\frac{1}{3}\left( \langle \delta^\textrm{QED} \rangle _\textrm{CT18} + \delta^\textrm{QED}_\textrm{MSHT20} + \delta^\textrm{QED}_\textrm{NNPDF3.1} \right), 
    \label{PDF4LHC21deltaQED}
\end{equation}
where the deltas on the right-hand side are the relative differences between the QCD+QED and QCD-only cross sections in Table~\ref{tab:xsecs}. The second row in Table~\ref{tab:xsecsPDF4LHC}
 shows the relative differences $\delta^\textrm{QED}_\textrm{PDF4LHC21}$ in percent, while the third row shows the central PDF4LHC21 cross sections after
correcting by $\delta^\textrm{QED}_\textrm{PDF4LHC21}$. The same QED corrections are to be applied to the predictions for PDF4LHC21 error PDFs when computing the PDF uncertainty.

\begin{table}[!htbp]
  \centering
  \begin{tabular}{l|l|l|l|l|l}
    \toprule
    PDF Set &
    $\sigma_{gg \to H}$ &
    $\sigma_{\rm VBF}$ &
    $\sigma_{Z H}$ &
    $\sigma_{W^+ H}$ &
    $\sigma_{W^- H}$ \\
    \midrule
    $\mbox{PDF4LHC21}_{40}$ NNLO      & 52.0 & 4.51 & 0.887 & 0.988 & 0.627 \\ \hline
    $\delta^\textrm{QED}_\textrm{PDF4LHC21}$ in \% & $-0.7$ & $-0.7$ & $-0.6$ & $-0.6$ & $-0.55$ \\
    $\mbox{PDF4LHC}_{\rm QED}$ NNLO   & 51.6 & 4.48 & 0.882 & 0.982 & 0.624 \\
    \bottomrule
  \end{tabular}
  \caption{Benchmark cross sections [pb] for various Higgs-production
           processes at $\sqrt{s}=14\;\text{TeV}$ obtained with PDF4LHC21\_40 NNLO PDFs, and with an applied QED correction explained in the main text.}
  \label{tab:xsecsPDF4LHC}
\end{table}

For CT18 NNLO, in Eq.~(\ref{PDF4LHC21deltaQED}) we average the corresponding three QED corrections in Table~\ref{tab:xsecs}, and for MSHT20 and NNPDF3.1, we take the respective NNLO QED corrections.  Such a prescription is sufficient for estimating the leading effects of the inclusion of QED effects into the PDFs without refitting the input ensembles of the PDF4LHC21 combination. Moreover, while this is necessarily an approximation,
our results have indicated that the PDF dependence of the QED corrections is typically small relative to their overall size. As such, our analysis indicates that the error induced by this approximation is notably smaller than the effect of not including the QED corrections at all. For example, when computing $\delta^\textrm{QED}_\textrm{PDF4LHC21}$ in Eq.~(\ref{PDF4LHC21deltaQED}), we could alternatively use the averages of MSHT20 NNLO and aN3LO, and NNPDF3.1 and 4.0 NNLO and aN3LO QED corrections -- this would not lead to appreciable changes in the QED corrections quoted in Table~\ref{tab:xsecsPDF4LHC} with the adopted prescription. Conversely, rather than averaging the CT18 QED corrections at NNLO, it would alternatively be acceptable to assume CT18$_{\rm QEDfit}$ in evaluating Eq.~(\ref{PDF4LHC21deltaQED}), leading to similar numerical results for $\delta^\textrm{QED}_\textrm{PDF4LHC21}$.

Understanding the origin of the observed differences between the PDF groups is work left to future studies. Clearly there are remaining uncertainties
arising from the implementation of the QED effects at the targeted level of precision. While a part of these uncertainties is 
captured by the input PDF analyses ({\it e.g.}, by propagating the QED correction from structure function data that constrain the photon PDF) and 
by our proposed simple rescaling procedure, another part could only be accounted for through a combination of the QED-enhanced PDF sets, which is beyond 
the scope of this work.
%

%
%
\section{Conclusion}
\label{sec:conc}
We have illustrated the size and behavior of QED effects on benchmark Higgs boson production at the LHC, propagated via QED-enhanced PDFs from three global analysis groups, CTEQ-TEA, MSHT, and NNPDF.  
We compared predictions based on the most up-to-date QED-enhanced PDF analyses that have
incorporated the LUX procedure for the photon PDF according to several possible approaches.
For total cross sections the spread over all predictions, independent of the inclusion of a QED PDF, is generally at the level of $\sim$few-percent variations, with the total rates for the processes explored in this study separated by $\sim\!3\!-\!4\%$ overall shifts from smallest to largest. Within this context, we find general consistency in the shifts induced by including QED-enhanced PDFs, which typically reduce total cross sections by $\sim\!\!0.5\!-\!1.5\%$ for the individual PDF sets before the explicit consideration of NLO electroweak effects, as explored in the specific case of Higgs-strahlung. These consistent reductions observed in the individual PDF sets, albeit of differing magnitudes, then result in reductions of $0.5\!-\!0.7\%$ in the average reweighting suggested in Section~\ref{sec:PDF4LHC}.
The few-percent PDF dependence of total Higgs cross sections carries through to unintegrated differential cross sections, as typified by $p^H_T$ spectra or charge-lepton rapidities in the Higgs-strahlung ($WH$) process with the leptonic decay of the $W$ boson, for which we also explored NLO electroweak corrections in parallel.
For this process, virtual NLO electroweak corrections typically diminish
total cross sections by $\sim\!8\%$, with this effect partially compensated by an opposing, positive
$\sim\!4\%$ contribution from photon-initiated graphs.
We quantified the size of this NLO electroweak photon-initiated effect by systematically including or excluding such contributions from the $W^+H$ process, finding a $\sim$few-percent shift at the scales for Higgs-strahlung observables computed in this study.
At the same time, we find that the QED PDF dependence of these NLO electroweak corrections themselves is very mild, falling overall at the $\sim$per-mille level.
Crucially, therefore, for precision goals in the Higgs sector, the widespread adoption of the LUX prescription for the photon PDF has substantially reduced the QED PDF-dependent spread in these NLO electroweak and photon-induced contributions, while
leaving an overall $\sim\!3\!-4\%$ residual PDF variation in the full, electroweak-corrected Higgs-strahlung cross sections, similar to the total cross section variations in other Higgs-production channels considered in our analysis.
The overwhelming source of this residual spread arises from differences in baseline QCD fits, rather than from the LUX structure-function input itself or associated PDF dependence in the NLO electroweak corrections.
Thus, we find evidence based on the Higgs-strahlung calculation that NLO electroweak corrections may often be assumed to have minimal PDF dependence, in much the same way that QED PDF effects in total cross sections may be estimated through averages like those shown in Sec.~\ref{sec:PDF4LHC}.
At the same time, we underscore that this conclusion applies specifically to the PDF dependence of the electroweak corrections where we have computed them (for Higgs-strahlung); the systematic and consistent inclusion of such effects for hadronic data fitted in global PDF fits remains an important challenge.

Going forward, the path to greater precision for Higgs physics therefore requires a higher level of control with respect to the base PDFs themselves, with less uncertainty attributable to ambiguities in the treatment of the photon PDF.
To this end, inter-group comparisons for benchmark channels involving Higgs production will play a valuable role.
As a crucial aspect of the LHC program is the realization of the most robust theoretical predictions in the Higgs sector and elsewhere, achieving higher precision in the PDFs, including with respect to QED effects, is a necessary task which will continue to drive work in this direction.

\vspace{0.1cm}

\section{Acknowledgments}
 We thank Juan Cruz Martinez for help to understand the QED effects in NNPDF. We also thank Lucian Harland-Lang, Robert Thorne, and PDF4LHC members for useful discussions. We thank members of the LHC Higgs Working Group for discussions that initiated this study.
 TC acknowledges funding by Research Foundation-Flanders (FWO) project number: 12E1323N. The work of TJH at Argonne National Laboratory was supported by the U.S.~Department of Energy under contract DE-AC02-06CH11357. The work of PMN is supported by the U.S.~Department of Energy under Grant No.~DE-SC0010129. PMN is grateful for support from the Wu-Ki Tung Endowed Chair in particle physics. KX is supported by the U.S. National Science Foundation under Grant No.~PHY-1607611 and PHY-2210452. This work of KX was performed in part at Aspen Center for Physics, which is supported by the U.S.~National Science Foundation under Grant No.~PHY-2210452.
 We gratefully acknowledge use of the Bebop supercomputer in the Laboratory Computing Resource Center at Argonne National Laboratory.

\newpage

\appendix

\section{Additional parton luminosities}
\label{app:lumis}

We collect several additional plots of the parton-parton luminosities here.

\begin{figure}[h!]
    \centering
    \includegraphics[width=6cm,height=3.750cm]{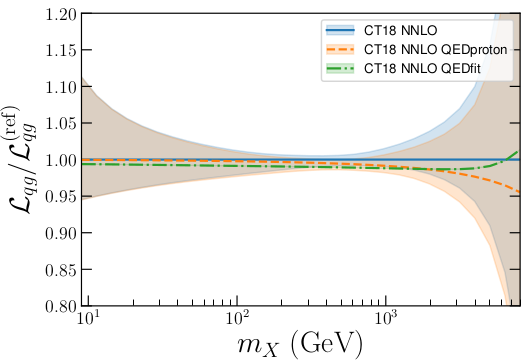}
    \includegraphics[width=6cm,height=3.750cm]{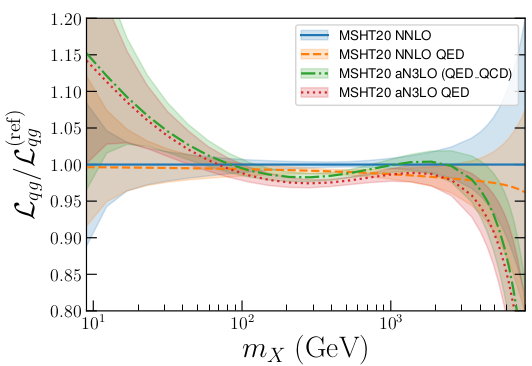}
    \includegraphics[width=6cm,height=3.750cm]{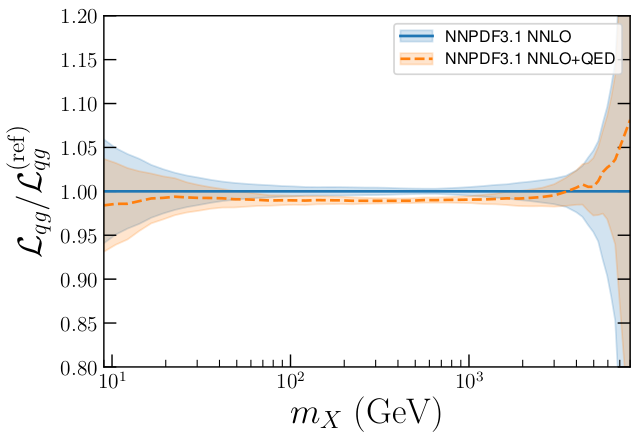} 
    \includegraphics[width=6cm,height=3.750cm]{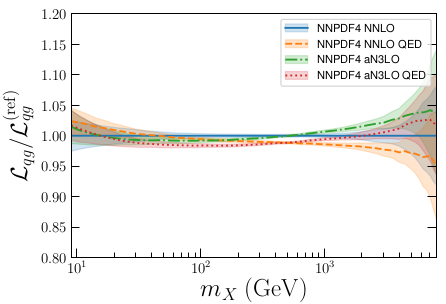}
    \vspace{-0.2cm}
    \caption{The quark-gluon parton luminosity for the CT18 (upper left), MSHT20 (upper right), NNPDF3.1 (lower left), and NNPDF4.0 without MHOU (lower right). The NNLO, NNLO+QED, and (where existing) aN3LO and aN3LO+QED PDFs are shown.}
    \label{fig:qg_PDF_lumis}
\end{figure}
\begin{figure}[h!]
    \centering
    \includegraphics[width=6cm,height=3.750cm]{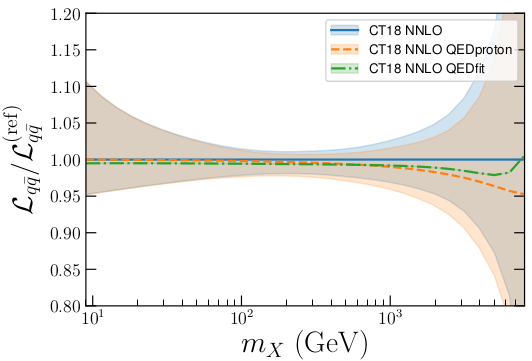}
    \includegraphics[width=6cm,height=3.750cm]{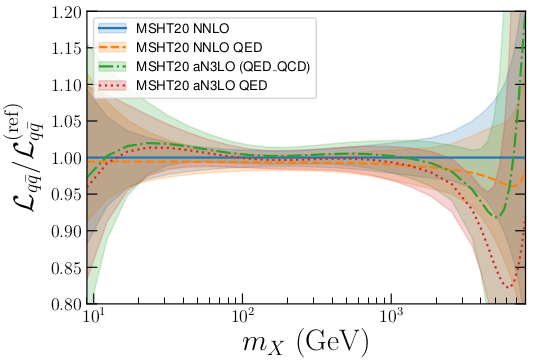}
    \includegraphics[width=6cm,height=3.750cm]{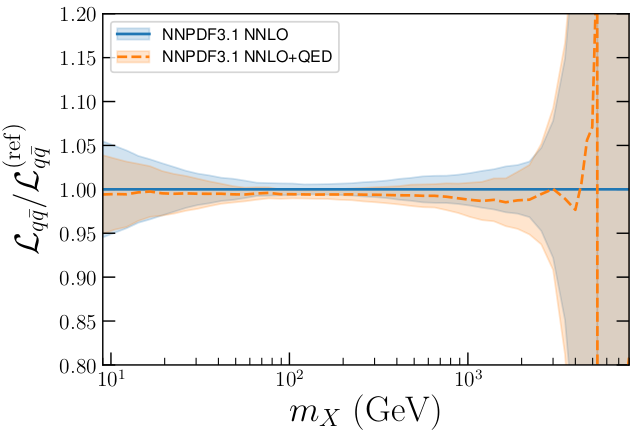} 
    \includegraphics[width=6cm,height=3.750cm]{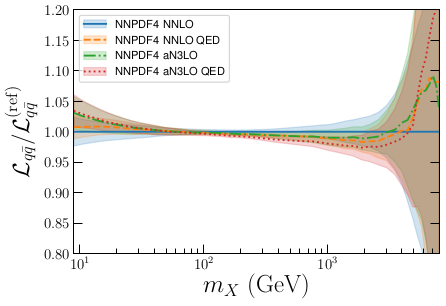}
    \vspace{-0.2cm}
    \caption{The quark-antiquark parton luminosity for the CT18 (upper left), MSHT20 (upper right), NNPDF3.1 (lower left), and NNPDF4.0 without MHOU (lower right). The NNLO, NNLO+QED, and (where existing) aN3LO and aN3LO+QED PDFs are shown.  In the case of NNPDF3.1, the quark-antiquark luminosities become negative at very large invariant masses, resulting in the spike seen in the ratio as it goes through 0.}
    \label{fig:qqbar_PDF_lumis}
\end{figure}

\newpage
%
%

\bibliographystyle{utphys}
\bibliography{qed-bench}

\providecommand{\href}[2]{#2}\begingroup\raggedright\begin{thebibliography}{10}

\bibitem{McGowan:2022nag}
J.~McGowan, T.~Cridge, L.~A. Harland-Lang, and R.~S. Thorne, ``{Approximate
  N$^{3}$LO parton distribution functions with theoretical uncertainties:
  MSHT20aN$^3$LO PDFs},''
  \href{http://dx.doi.org/10.1140/epjc/s10052-023-11236-0}{{\em Eur. Phys. J.
  C} {\bfseries 83} no.~3, (2023) 185},
  \href{http://arxiv.org/abs/2207.04739}{{\ttfamily arXiv:2207.04739
  [hep-ph]}}. [Erratum: Eur.Phys.J.C 83, 302 (2023)].

\bibitem{Jing:2023isu}
X.~Jing {\em et~al.}, ``{Quantifying the interplay of experimental constraints
  in analyses of parton distributions},''
  \href{http://dx.doi.org/10.1103/PhysRevD.108.034029}{{\em Phys. Rev. D}
  {\bfseries 108} no.~3, (2023) 034029},
  \href{http://arxiv.org/abs/2306.03918}{{\ttfamily arXiv:2306.03918
  [hep-ph]}}.

\bibitem{Cridge:2023ozx}
T.~Cridge, L.~A. Harland-Lang, and R.~S. Thorne, ``{The impact of LHC jet and
  Z$p_T$ data at up to approximate N${}^3$LO order in the MSHT global PDF
  fit},'' \href{http://dx.doi.org/10.1140/epjc/s10052-024-12771-0}{{\em Eur.
  Phys. J. C} {\bfseries 84} no.~4, (2024) 446},
  \href{http://arxiv.org/abs/2312.12505}{{\ttfamily arXiv:2312.12505
  [hep-ph]}}.

\bibitem{Andersen:2024czj}
J.~Andersen {\em et~al.}, ``{Les Houches 2023: Physics at TeV Colliders:
  Standard Model Working Group Report},'' in {\em {Physics of the TeV Scale and
  Beyond the Standard Model}: {Intensifying the Quest for New Physics}}.
\newblock 6, 2024.
\newblock \href{http://arxiv.org/abs/2406.00708}{{\ttfamily arXiv:2406.00708
  [hep-ph]}}.

\bibitem{NNPDF:2024nan}
{\bfseries NNPDF} Collaboration, R.~D. Ball {\em et~al.}, ``{The path to $\hbox
  {N}^3\hbox {LO}$ parton distributions},''
  \href{http://dx.doi.org/10.1140/epjc/s10052-024-12891-7}{{\em Eur. Phys. J.
  C} {\bfseries 84} no.~7, (2024) 659},
  \href{http://arxiv.org/abs/2402.18635}{{\ttfamily arXiv:2402.18635
  [hep-ph]}}.

\bibitem{Cooper-Sarkar:2024crx}
A.~Cooper-Sarkar, T.~Cridge, F.~Giuli, L.~A. Harland-Lang, F.~Hekhorn,
  J.~Huston, G.~Magni, S.~Moch, and R.~S. Thorne, ``{A Benchmarking of QCD
  Evolution at Approximate $N^3LO$},''
  \href{http://arxiv.org/abs/2406.16188}{{\ttfamily arXiv:2406.16188
  [hep-ph]}}.

\bibitem{MSHT:2024tdn}
{\bfseries MSHT, NNPDF} Collaboration, T.~Cridge {\em et~al.}, ``{Combination
  of aN$^3$LO PDFs and implications for Higgs production cross-sections at the
  LHC},'' \href{http://arxiv.org/abs/2411.05373}{{\ttfamily arXiv:2411.05373
  [hep-ph]}}.

\bibitem{Cridge:2023ryv}
T.~Cridge, L.~A. Harland-Lang, and R.~S. Thorne, ``{Combining QED and
  approximate ${\rm N}^3$LO QCD corrections in a global PDF fit:
  MSHT20qed\_an3lo PDFs},''
  \href{http://dx.doi.org/10.21468/SciPostPhys.17.1.026}{{\em SciPost Phys.}
  {\bfseries 17} no.~1, (2024) 026},
  \href{http://arxiv.org/abs/2312.07665}{{\ttfamily arXiv:2312.07665
  [hep-ph]}}.

\bibitem{Barontini:2024dyb}
A.~Barontini, N.~Laurenti, and J.~Rojo, ``{NNPDF4.0 aN$^3$LO PDFs with QED
  corrections},'' in {\em {31st International Workshop on Deep-Inelastic
  Scattering and Related Subjects}}.
\newblock 6, 2024.
\newblock \href{http://arxiv.org/abs/2406.01779}{{\ttfamily arXiv:2406.01779
  [hep-ph]}}.

\bibitem{Manohar:2016nzj}
A.~Manohar, P.~Nason, G.~P. Salam, and G.~Zanderighi, ``{How bright is the
  proton? A precise determination of the photon parton distribution
  function},'' \href{http://dx.doi.org/10.1103/PhysRevLett.117.242002}{{\em
  Phys. Rev. Lett.} {\bfseries 117} no.~24, (2016) 242002},
  \href{http://arxiv.org/abs/1607.04266}{{\ttfamily arXiv:1607.04266
  [hep-ph]}}.

\bibitem{Manohar:2017eqh}
A.~V. Manohar, P.~Nason, G.~P. Salam, and G.~Zanderighi, ``{The Photon Content
  of the Proton},'' \href{http://dx.doi.org/10.1007/JHEP12(2017)046}{{\em JHEP}
  {\bfseries 12} (2017) 046}, \href{http://arxiv.org/abs/1708.01256}{{\ttfamily
  arXiv:1708.01256 [hep-ph]}}.

\bibitem{Xie:2021equ}
{\bfseries CTEQ-TEA} Collaboration, K.~Xie, T.~J. Hobbs, T.-J. Hou, C.~Schmidt,
  M.~Yan, and C.~P. Yuan, ``{Photon PDF within the CT18 global analysis},''
  \href{http://dx.doi.org/10.1103/PhysRevD.105.054006}{{\em Phys. Rev. D}
  {\bfseries 105} no.~5, (2022) 054006},
  \href{http://arxiv.org/abs/2106.10299}{{\ttfamily arXiv:2106.10299
  [hep-ph]}}.

\bibitem{Xie:2023qbn}
{\bfseries CTEQ-TEA} Collaboration, K.~Xie, B.~Zhou, and T.~J. Hobbs, ``{The
  photon content of the neutron},''
  \href{http://dx.doi.org/10.1007/JHEP04(2024)022}{{\em JHEP} {\bfseries 04}
  (2024) 022}, \href{http://arxiv.org/abs/2305.10497}{{\ttfamily
  arXiv:2305.10497 [hep-ph]}}.

\bibitem{Cridge:2021pxm}
T.~Cridge, L.~A. Harland-Lang, A.~D. Martin, and R.~S. Thorne, ``{QED parton
  distribution functions in the MSHT20 fit},''
  \href{http://dx.doi.org/10.1140/epjc/s10052-022-10028-2}{{\em Eur. Phys. J.
  C} {\bfseries 82} no.~1, (2022) 90},
  \href{http://arxiv.org/abs/2111.05357}{{\ttfamily arXiv:2111.05357
  [hep-ph]}}.

\bibitem{NNPDF:2024djq}
{\bfseries NNPDF} Collaboration, R.~D. Ball {\em et~al.}, ``{Photons in the
  proton: implications for the LHC},''
  \href{http://dx.doi.org/10.1140/epjc/s10052-024-12731-8}{{\em Eur. Phys. J.
  C} {\bfseries 84} no.~5, (2024) 540},
  \href{http://arxiv.org/abs/2401.08749}{{\ttfamily arXiv:2401.08749
  [hep-ph]}}.

\bibitem{Barontini:2024eii}
A.~Barontini, N.~Laurenti, and J.~Rojo, ``{NNPDF progress and the path to
  proton structure at N$^3$LO accuracy},''
  \href{http://dx.doi.org/10.22323/1.469.0039}{{\em PoS} {\bfseries DIS2024}
  (2025) 039}.

\bibitem{Amoroso:2022eow}
S.~Amoroso {\em et~al.}, ``{Snowmass 2021 Whitepaper: Proton Structure at the
  Precision Frontier},''
  \href{http://dx.doi.org/10.5506/APhysPolB.53.12-A1}{{\em Acta Phys. Polon. B}
  {\bfseries 53} no.~12, (2022) 12--A1},
  \href{http://arxiv.org/abs/2203.13923}{{\ttfamily arXiv:2203.13923
  [hep-ph]}}.

\bibitem{PDF4LHCWorkingGroup:2022cjn}
{\bfseries PDF4LHC Working Group} Collaboration, R.~D. Ball {\em et~al.},
  ``{The PDF4LHC21 combination of global PDF fits for the LHC Run III},''
  \href{http://dx.doi.org/10.1088/1361-6471/ac7216}{{\em J. Phys. G} {\bfseries
  49} no.~8, (2022) 080501}, \href{http://arxiv.org/abs/2203.05506}{{\ttfamily
  arXiv:2203.05506 [hep-ph]}}.

\bibitem{Cridge:2021qjj}
{\bfseries PDF4LHC21 combination group} Collaboration, T.~Cridge, ``{PDF4LHC21:
  Update on the benchmarking of the CT, MSHT and NNPDF global PDF fits},''
  \href{http://dx.doi.org/10.21468/SciPostPhysProc.8.101}{{\em SciPost Phys.
  Proc.} {\bfseries 8} (2022) 101},
  \href{http://arxiv.org/abs/2108.09099}{{\ttfamily arXiv:2108.09099
  [hep-ph]}}.

\bibitem{Harland-Lang:2019pla}
L.~A. Harland-Lang, A.~D. Martin, R.~Nathvani, and R.~S. Thorne, ``{Ad Lucem:
  QED Parton Distribution Functions in the MMHT Framework},''
  \href{http://dx.doi.org/10.1140/epjc/s10052-019-7296-0}{{\em Eur. Phys. J. C}
  {\bfseries 79} no.~10, (2019) 811},
  \href{http://arxiv.org/abs/1907.02750}{{\ttfamily arXiv:1907.02750
  [hep-ph]}}.

\bibitem{Hou:2019efy}
T.-J. Hou {\em et~al.}, ``{New CTEQ global analysis of quantum chromodynamics
  with high-precision data from the LHC},''
  \href{http://dx.doi.org/10.1103/PhysRevD.103.014013}{{\em Phys. Rev. D}
  {\bfseries 103} no.~1, (2021) 014013},
  \href{http://arxiv.org/abs/1912.10053}{{\ttfamily arXiv:1912.10053
  [hep-ph]}}.

\bibitem{Bailey:2020ooq}
S.~Bailey, T.~Cridge, L.~A. Harland-Lang, A.~D. Martin, and R.~S. Thorne,
  ``{Parton distributions from LHC, HERA, Tevatron and fixed target data:
  MSHT20 PDFs},'' \href{http://dx.doi.org/10.1140/epjc/s10052-021-09057-0}{{\em
  Eur. Phys. J. C} {\bfseries 81} no.~4, (2021) 341},
  \href{http://arxiv.org/abs/2012.04684}{{\ttfamily arXiv:2012.04684
  [hep-ph]}}.

\bibitem{Buckley:2014ana}
A.~Buckley, J.~Ferrando, S.~Lloyd, K.~Nordstr{\"o}m, B.~Page, M.~R{\"u}fenacht,
  M.~Sch{\"o}nherr, and G.~Watt, ``{LHAPDF6: parton density access in the LHC
  precision era},''
  \href{http://dx.doi.org/10.1140/epjc/s10052-015-3318-8}{{\em Eur. Phys. J. C}
  {\bfseries 75} (2015) 132}, \href{http://arxiv.org/abs/1412.7420}{{\ttfamily
  arXiv:1412.7420 [hep-ph]}}.

\bibitem{NNPDF:2017mvq}
{\bfseries NNPDF} Collaboration, R.~D. Ball {\em et~al.}, ``{Parton
  distributions from high-precision collider data},''
  \href{http://dx.doi.org/10.1140/epjc/s10052-017-5199-5}{{\em Eur. Phys. J. C}
  {\bfseries 77} no.~10, (2017) 663},
  \href{http://arxiv.org/abs/1706.00428}{{\ttfamily arXiv:1706.00428
  [hep-ph]}}.

\bibitem{NNPDF:2021njg}
{\bfseries NNPDF} Collaboration, R.~D. Ball {\em et~al.}, ``{The path to proton
  structure at 1\% accuracy},''
  \href{http://dx.doi.org/10.1140/epjc/s10052-022-10328-7}{{\em Eur. Phys. J.
  C} {\bfseries 82} no.~5, (2022) 428},
  \href{http://arxiv.org/abs/2109.02653}{{\ttfamily arXiv:2109.02653
  [hep-ph]}}.

\bibitem{NNPDF4QED_Website}
NNPDF4.0 QED Website:
  \url{https://nnpdf.mi.infn.it/wp-content/uploads/2024/05/NNPDF40\_nnlo\_as\_01180\_qcd.tar.gz}.

\bibitem{Bertone:2017bme}
{\bfseries NNPDF} Collaboration, V.~Bertone, S.~Carrazza, N.~P. Hartland, and
  J.~Rojo, ``{Illuminating the photon content of the proton within a global PDF
  analysis},'' \href{http://dx.doi.org/10.21468/SciPostPhys.5.1.008}{{\em
  SciPost Phys.} {\bfseries 5} no.~1, (2018) 008},
  \href{http://arxiv.org/abs/1712.07053}{{\ttfamily arXiv:1712.07053
  [hep-ph]}}.

\bibitem{Baglio:2022wzu}
J.~Baglio, C.~Duhr, B.~Mistlberger, and R.~Szafron, ``{Inclusive production
  cross sections at N$^{3}$LO},''
  \href{http://dx.doi.org/10.1007/JHEP12(2022)066}{{\em JHEP} {\bfseries 12}
  (2022) 066}, \href{http://arxiv.org/abs/2209.06138}{{\ttfamily
  arXiv:2209.06138 [hep-ph]}}.

\bibitem{Dreyer:2018qbw}
F.~A. Dreyer and A.~Karlberg, ``{Vector-Boson Fusion Higgs Pair Production at
  N$^3$LO},'' \href{http://dx.doi.org/10.1103/PhysRevD.98.114016}{{\em Phys.
  Rev. D} {\bfseries 98} no.~11, (2018) 114016},
  \href{http://arxiv.org/abs/1811.07906}{{\ttfamily arXiv:1811.07906
  [hep-ph]}}.

\bibitem{Denner:2014cla}
A.~Denner, S.~Dittmaier, S.~Kallweit, and A.~M\"uck, ``{HAWK 2.0: A Monte Carlo
  program for Higgs production in vector-boson fusion and Higgs strahlung at
  hadron colliders},'' \href{http://dx.doi.org/10.1016/j.cpc.2015.04.021}{{\em
  Comput. Phys. Commun.} {\bfseries 195} (2015) 161--171},
  \href{http://arxiv.org/abs/1412.5390}{{\ttfamily arXiv:1412.5390 [hep-ph]}}.

\bibitem{LHCHiggsCrossSectionWorkingGroup:2016ypw}
{\bfseries LHC Higgs Cross Section Working Group} Collaboration, D.~de~Florian
  {\em et~al.}, ``{Handbook of LHC Higgs Cross Sections: 4. Deciphering the
  Nature of the Higgs Sector},''
  \href{http://arxiv.org/abs/1610.07922}{{\ttfamily arXiv:1610.07922
  [hep-ph]}}.

\bibitem{Spira:2016ztx}
M.~Spira, ``{Higgs Boson Production and Decay at Hadron Colliders},''
  \href{http://dx.doi.org/10.1016/j.ppnp.2017.04.001}{{\em Prog. Part. Nucl.
  Phys.} {\bfseries 95} (2017) 98--159},
  \href{http://arxiv.org/abs/1612.07651}{{\ttfamily arXiv:1612.07651
  [hep-ph]}}.

\bibitem{Jones:2023uzh}
S.~P. Jones, ``{An Overview of Standard Model Calculations for Higgs Boson
  Production \& Decay},'' \href{http://dx.doi.org/10.31526/lhep.2023.442}{{\em
  LHEP} {\bfseries 2023} (2023) 442}.

\bibitem{Karlberg:2024zxx}
A.~Karlberg {\em et~al.}, ``{Ad interim recommendations for the Higgs boson
  production cross sections at $\sqrt{s} = 13.6$ TeV},''
  \href{http://arxiv.org/abs/2402.09955}{{\ttfamily arXiv:2402.09955
  [hep-ph]}}.

\bibitem{LesHouches2025}
T.~Cridge and J.~M. Cruz-Martinez, ``{Les Houches 2025 Proceedings},''
\newblock 2025.
\newblock \href{http://arxiv.org/abs/25xx.xxxxx}{{\ttfamily arXiv:25xx.xxxxx
  [hep-ph]}}.

\bibitem{Maltoni:2017ims}
F.~Maltoni, D.~Pagani, A.~Shivaji, and X.~Zhao, ``{Trilinear Higgs coupling
  determination via single-Higgs differential measurements at the LHC},''
  \href{http://dx.doi.org/10.1140/epjc/s10052-017-5410-8}{{\em Eur. Phys. J. C}
  {\bfseries 77} no.~12, (2017) 887},
  \href{http://arxiv.org/abs/1709.08649}{{\ttfamily arXiv:1709.08649
  [hep-ph]}}.

\bibitem{Obul:2018psx}
P.~Obul, S.~Dulat, T.-J. Hou, A.~Tursun, and N.~Yalkun, ``{Next-to-leading
  order QCD and electroweak corrections to Higgs-strahlung processes at the
  LHC},'' \href{http://dx.doi.org/10.1088/1674-1137/42/9/093105}{{\em Chin.
  Phys. C} {\bfseries 42} no.~9, (2018) 093105},
  \href{http://arxiv.org/abs/1801.06851}{{\ttfamily arXiv:1801.06851
  [hep-ph]}}.

\end{thebibliography}\endgroup

\end{document}